\documentclass[12pt]{article}
\usepackage{amsmath}
\usepackage{amssymb}
\usepackage{amsfonts}
\usepackage{graphicx}
\usepackage[utf8]{inputenc}

\usepackage{slashed}       
\usepackage{mathrsfs}      
\usepackage{bbm}           

\usepackage[dvipsnames]{xcolor}
\usepackage{fancyhdr}		
\usepackage{lipsum}         
\usepackage{framed}			
\usepackage{cite}          

\usepackage{booktabs}		
\usepackage{nicefrac}		

\usepackage[font=small]{caption}
\usepackage{float}         
\usepackage{subcaption}

\usepackage[margin=2cm]{geometry}
\graphicspath{{figures/}}
\numberwithin{equation}{section}
\usepackage{tocloft}
\usepackage{multicol}
\usepackage{etoolbox}
\usepackage{relsize}
\patchcmd{\thebibliography}
  {\list}
  {\begin{multicols}{2}\smaller\list}
  {}
  {}
\appto{\endthebibliography}{\end{multicols}}

\let\oldenumerate\enumerate
\renewcommand{\enumerate}{
  \oldenumerate
  \setlength{\itemsep}{1pt}
  \setlength{\parskip}{0pt}
  \setlength{\parsep}{0pt}
}

\let\olditemize\itemize
\renewcommand{\itemize}{
  \olditemize
  \setlength{\itemsep}{1pt}
  \setlength{\parskip}{0pt}
  \setlength{\parsep}{0pt}
}

\newcommand{\acro}[1]{\textsc{\MakeLowercase{#1}}}

\renewcommand{\tilde}{\widetilde}   

\DeclareMathOperator{\tr}{Tr}
\DeclareMathOperator{\Det}{Det}

\newcommand{\SU}[1]{\text{\acro{SU}{\footnotesize#1}}}

\newcommand{\UU}[1]{\text{\acro{U}{\footnotesize#1}}}

\newcommand{\email}[1]{\href{mailto:#1}{#1}}
\newenvironment{institutions}[1][2em]{\begin{list}{}{\setlength\leftmargin{#1}\setlength\rightmargin{#1}}\item[]}{\end{list}}

\usepackage[
	colorlinks=true,
	citecolor= green!50!black,
	linkcolor= NavyBlue!75!black,
	urlcolor=green!50!black,
	hypertexnames=false]{hyperref}

\fancypagestyle{firststyle}{
	\rhead{\footnotesize%
	\texttt{UCR-TR-FLIP-2019-NX-74205}
	}}

\usepackage{etoc}

\begin{document}

\begin{center}

    {\huge \bf Vector Self-Interacting Dark Matter}

    \vskip .7cm

   { \bf
    Ian Chaffey
    and
    Philip Tanedo}
    \\ \vspace{-.2em}
    { \tt
    \footnotesize
    \email{ichaf001@ucr.edu},
    \email{flip.tanedo@ucr.edu}
    }

    \vspace{-.2cm}

    \begin{institutions}[2.25cm]
    \footnotesize
	{\it
	    Department of Physics \& Astronomy,
	    University of  California, Riverside,
	    {CA} 92521
	    }
    \end{institutions}

\end{center}


\begin{abstract}
\noindent
We present a model of vector dark matter that interacts through a low-mass vector mediator based on the Higgsing of an \SU{(2)} dark sector. The dark matter is charged under a \UU{(1)} gauge symmetry. Even though this symmetry is broken, the residual global symmetries of the theory prevent dark matter decay. We present the behavior of the model subject to the assumption that the dark matter abundance is due to thermal freeze out, including self-interaction targets for small scale structure anomalies and the possibility of interacting with the Standard Model through the vector mediator.
\end{abstract}

\small
\setcounter{tocdepth}{2}
\footnotesize
\setcounter{tocdepth}{2}
\tableofcontents
\normalsize
\newpage
\normalsize


\section{Introduction and Context}

Despite strong evidence for the existence of dark matter~\cite{Ade:2015xua,Aghanim:2018eyx}, the lack of a definitive signal in recent experiments puts pressure on the well-studied weakly-interacting massive particle (\acro{WIMP}) paradigm~\cite{Tan:2016zwf, Akerib:2015rjg, Akerib:2016vxi,
Agnese:2015nto, Hooper:2012sr, Ackermann:2015zua, Bergstrom:2013jra, Cirelli:2013hv}. One approach beyond this framework is to assume that dark matter belongs to a decoupled sector of particles frequently referred to as dark or hidden sectors with low-mass particles that mediate interactions~\cite{%
Pospelov:2008zw,
Pospelov:2008jd,
Pospelov:2007mp,
Essig:2013lka,
Alexander:2016aln,
Battaglieri:2017aum
}.
A simple realization of this is the dark photon portal in which a low-mass spin-1 vector boson couples to the Standard Model through kinetic mixing with the hypercharge gauge bosons~\cite{Holdom:1985ag, GALISON1984279}. Instead of annihilating directly into Standard Model particles, dark matter annihilates into dark photons that subsequently decay into Standard Model particles. Dark photons could be detected by a number of current and future experiments~\cite{Essig:2013lka, Alexander:2016aln, Battaglieri:2017aum}.
An automatic feature of dark sector models is the existence of long-range, velocity-dependent self-interactions between dark matter particles coming from exchange of the low-mass mediator. These self-interactions between dark matter particles can address several potential small scale structure tensions between simulations of cold dark matter and astronomical observations~\cite{Tulin:2017ara}.

This manuscript introduces a model of spin-1 dark matter that self-interacts through low-mass, spin-1 mediators (dark photons). The dark sector is composed of a \SU{(2)} gauge group with a scalar sector that enacts two stages of symmetry breaking:
\begin{enumerate}
	\item \SU{(2)}$\to$\UU{(1)} at a scale $f$, which sets the scale of the dark matter particles, and
	\item \SU{(2)}$\to \varnothing$ at a scale $v \ll f$, which sets the scale of the dark radiation.
\end{enumerate}
We appeal to the analogy of massive $W^\pm$ bosons interacting with a \emph{massive} photon, a structure that is similar to the  ordinary electroweak sector. The stability of the dark matter is ensured by a residual global \UU{(1)} in the theory. This is the first spin-1 dark sector theory with a massive spin-1 mediator coming from the same multiplet as the dark matter.


Compared to fermionic or spin-0 candidates, vector bosons are a relatively unexplored dark matter candidate~\cite{Servant:2002aq,
Hambye:2008bq, Farzan:2012hh, Davoudiasl:2013jma, Gross:2015cwa, Ko:2016fcd,Baek:2013dwa, Khoze:2014woa,Boddy:2014qxa,Boddy:2014yra,
Elahi:2019jeo, Choi:2019zeb}.
The first proposal of spin-1 dark matter was the Kaluza--Klein photon in the universal extra dimension scenario. This is a spin-1 analog to the supersymmetric neutralino: it is a weakly-interacting massive particle whose existence is related to a symmetry solution of the Higgs hierarchy problem. 5D translation invariance ensures dark matter stability~\cite{Servant:2002aq}, though the scenario is constrained by collider searches because the visible matter fields also extend into the extra dimension~\cite{Deutschmann:2017bth}.

Non-universal extra dimensional scenarios may avoid collider bounds, but typically require additional features to stabilize dark matter from decaying.
Later models explored non-Abelian spin-1 dark mater purely in a hidden sector; these dark sector constructions differ from typical weakly-interacting massive particles in that they do not begin with the assumption that the new particles are related to the naturalness of the Standard Model Higgs sector~\cite{Alexander:2016aln}.
Simple constructions with a $\SU{(2)}$ gauge group provide degenerate, massive spin-1 particles that can be stable due to custodial symmetry~\cite{Hambye:2008bq, Davoudiasl:2013jma, Gross:2015cwa}. 
Other models are based on the spontaneous breaking of scale invariance~\cite{YaserAyazi:2019caf,Baldes:2018emh,Karam:2016rsz,Khoze:2016zfi,Karam:2015jta,Khoze:2014xha,Carone:2013wla,Hambye:2013sna}.
The scalar field that breaks the gauge symmetry may be used as a portal to the visible sector by mixing with the Standard Model Higgs; the amount of mixing controls the signal at direct detection experiments. A recent exploration with $\SU{(3)}$ gauge group may resolve a tension in the Hubble constant measurement~\cite{Ko:2016fcd}. Our study focuses a scenario where the triplet of $\SU{(2)}$ gauge bosons separate into hidden-charged dark matter (analogs of the $W^\pm$) and a massive dark photon (analog of the $A$), which may then kinetically mix with the visible sector photon.

$\SU{(2)}$ sectors admit monopoles when there is an unbroken $\UU{(1)}$ subgroup. This leads to studies of dark sectors that contain both vector dark matter and dark 't Hooft--Polyakov monopoles~\cite{Baek:2013dwa, Khoze:2014woa}. This phenomena becomes more subtle in the case we study because the $\UU{(1)}$ global symmetry is Higgsed so that the monopoles confine. We leave a study of this case for future work.
An orthogonal direction in the study of non-Abelian dark sectors is the case where the gauge theory confines. In this phase one has strongly-interacting dark matter composed of glueball-like states~\cite{Boddy:2014qxa,Boddy:2014yra}. Our model differs from this in that it is Higgsed rather than confined, allowing the dark matter states to be massive spin-1 particles. Alternatively, Ref.~\cite{Choi:2019zeb} recently studied vector strongly interacting dark matter. Our model differs in that it is a simple gauge group with a different scalar content and standard dark sector annihilation modes.

\section{Particles and Symmetries}

An $\SU{(2)}$ gauge field $W^a_{\mu}$ couples with strength $g$ to a two scalar particles: a doublet $H^i$ and an adjoint scalar $\Phi=\phi^a T^a$. In this representation, the $\SU{(2)}$ transformation is
\begin{align}
  H(x) &\to U H(x)
  &
  \Phi(x)  &\to  U \Phi(x) U^\dag \ ,
  \label{eq:H:Phi:transform}
\end{align}
where $U = \text{exp}(i\theta^aT^a)$ is a $2\times 2$ special unitary matrix and $T^a = \frac 12 \sigma^a$ are the generators of $\SU{(2)}$ in the fundamental representation.
In the limit of no interactions, the particles respect a global ``flavor'' symmetry
\begin{align}
\SU{(2)}_\Phi \times \SU{(2)}_H \times \UU{(1)}_H
\;=\;
\SU{(2)}_\text{V} \times \SU{(2)}_\text{A} \times \UU{(1)}_H \ ,
\end{align}
under which the scalar fields transform as
\begin{align}
\SU{(2)}_\Phi: &\; \Phi \to  U_\Phi \Phi  U_\Phi^\dag
&
\SU{(2)}_H: &\; H \to  U_H H
&
\UU{(1)}_H: &\; H \to e^{i\theta_H} H \ .
\end{align}
We gauge the diagonal (vector) subgroup $\SU{(2)}_\text{V}$ of $\SU{(2)}_\Phi \times \SU{(2)}_H$ composed of transformations with $U_\Phi = U_H$. The orthogonal combination is the axial symmetry $\SU{(2)}_\text{A}$, for which $U_\Phi = U_H^\dag$. The $\UU{(1)}_H$ ``Higgs number'' symmetry is analogous to hypercharge in the Standard Model.

\subsection{General, Renormalizable Lagrangian}

The general, renormalizable Lagrangian satisfying the global symmetries of the particle content is
\begin{align}
\mathcal L &=
-\frac{1}{4} F^a_{\mu \nu} F^{a \mu \nu}
+ \left|D_\mu H\right|^2
+ \tr |\mathcal{D}_{\mu}\Phi|^2
- V
\label{eq:L}
\\
V & =
\phantom{+}
\frac{\lambda}{4!}
\left(
  2 \tr \Phi^2 - f_0^2
\right)^2
+
\frac{\lambda'}{4!}
\left(
  2|H|^2 - v_0^2
\right)^2
+ \mu H^{\dagger}\Phi H
+ \lambda'' |H|^2 \tr \Phi^2
\label{eq:V} \ .
\end{align}
$D$ and $\mathcal D$ are covariant derivatives for the fundamental and adjoint of $\SU{(2)}$, respectively.
We write the potential $V$ to imply that the scalars $\Phi$ and $H$ obtain vacuum expectation values (vevs) that spontaneously break the symmetries of the theory. This breaking produces a spectrum of Goldstone bosons, three of which are eaten by the massive gauge bosons.
The trilinear $\mu$ term explicitly breaks the global axial $\SU{(2)}_\text{A}$ symmetry. This gives a mass to the remaining the would-be Goldstone modes. The $\lambda''$ term mixes the radial modes of the $H$ and $\Phi$. We systematically examine the theory starting from the symmetry breaking $\lambda$ and $\lambda'$ terms and subsequently include the effects of the $\mu$ and $\lambda''$ terms.
Additional quartic terms obeying the global symmetries reduce to the $\lambda''$ term.\footnote{For example:
$ \displaystyle
  H^\dag \Phi^2 H = \frac 12 H^\dag \left\{\Phi, \Phi\right\} H = \frac{1}{2}H^\dag \left( \frac 12 \phi^a \phi^b \delta^{ab}\mathbf{1}_{2\times 2}\right) H
  = \frac 12 |H|^2 \tr \Phi^2\ .
$
}

One may also consider additional potential terms that use the pseudo-conjugate field $\tilde H^i \equiv \epsilon^{ij}H^\dag_{\phantom\dag j}$, exploiting the pseudoreality of $\SU{(2)}$. Analogously to the Standard Model, $\tilde H$ transforms like $H$ with respect to $\SU{(2)}_H$ but with opposite charge under $\UU{(1)}_H$. Any renormalizable potential terms written with $\tilde H$ either reduce to terms in \eqref{eq:V} or explicitly violate the $\UU{(1)}_H$ symmetry. We assume the case that this Higgs number symmetry is respected at the Lagrangian level and so we do not include any such terms.

\subsection{Spectrum, Symmetry, Stability}

A qualitative overview of the model is as follows. The vacuum of the scalar potential spontaneously breaks the global symmetry $\SU{(2)}_\Phi \times \SU{(2)}_H \times \UU{(1)}_H \to \UU{(1)}_{H'}$, where $\UU{(1)}_{H'}$ is generated by
\begin{align}
\UU{(1)}_{H'}:\qquad&  T^3_V + \frac 12 T_H \, ,
\end{align}
analogous to electric charge in the electroweak sector. In what follows, we refer to the \emph{charge} of a dark sector particle with respect to the $\UU{(1)}_\text{V} \subset \SU{(2)}_\text{V}$ gauge symmetry of the mediator. The gauge bosons eat three of the five Goldstone modes. We suggestively name the remaining two `pions,' $\pi^\pm$.
We take the limit where the triplet vev is much larger than the doublet vev,
\begin{align}
  \langle\text{Tr}\,\Phi^2\rangle = \frac{f^2}{2}
  \qquad\gg\qquad
  \langle |H|^2 \rangle = \frac{v^2}{2} \ .
\end{align}
Then the particle content in the $\mu = \lambda'' = 0$ limit are:
\begin{enumerate}
  \item \textbf{Dark matter}: $W^\pm$ gauge bosons with mass $\sim gf$; primarily eats the Goldstones in the $\Phi$.
  \item \textbf{Mediator}: $A$ gauge boson with mass $\sim gv$, eats the neutral Goldstone in $H$.
  \item \textbf{Dark pions}: $\pi^\pm$ charged scalars that are mostly the charged Goldstones in $H$.
\end{enumerate}
We write $W^\pm$, $A$, $H$, and $\pi^\pm$ to suggest parallels to the Standard Model electroweak gauge fields, Higgs, and charged pions. However, our fields are completely distinct from their visible sector counterparts. For example, there is no $Z$ boson analog since only $\SU{(2)}_\text{V}$ is gauged.

\begin{figure}
  \begin{center}
    \includegraphics[width=\textwidth]{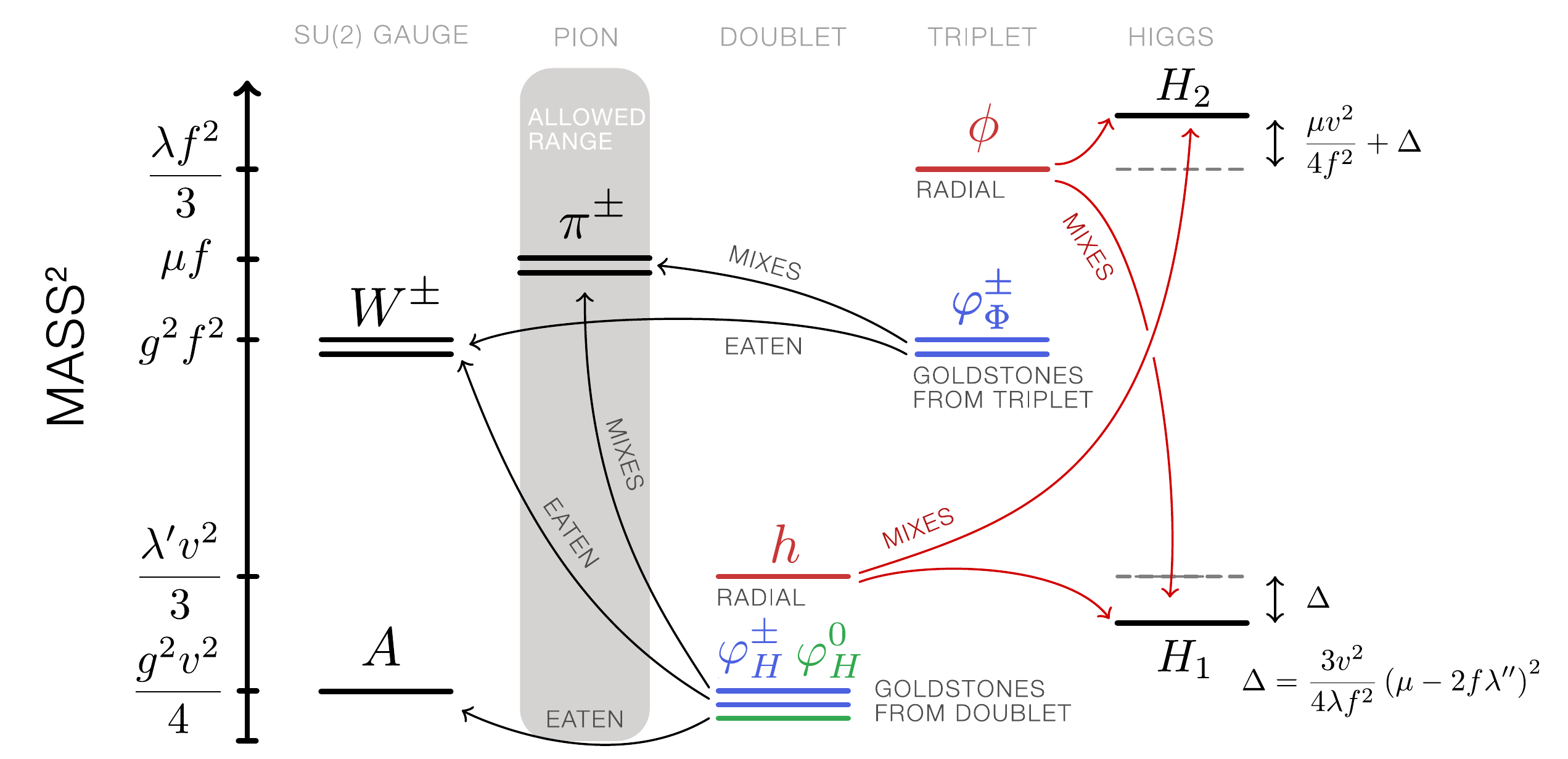}
  \end{center}
  \caption{Model spectrum. Mass eigenstates are black lines, charged (neutral) Goldstones are blue (green) lines, radial Higgs modes are red lines. Mixing into mass eigenstates indicated by thin lines.}
  \label{fig:spectrum}
\end{figure}

The key features of this model are:
\begin{itemize}
  \item The $W^\pm$ and $\pi^\pm$ are labeled with respect to their charge with respect to $A$. However, this charge is not conserved due to the doublet vev $\langle H \rangle$. It cannot be used to stabilize the dark matter. This is a key difference from other $\SU{(2)}\to \UU{(1)}$ models of vector dark matter~\cite{Baek:2013dwa}.
  \item The stability of $W^\pm$ is enforced by (1) the unbroken $\UU{(1)}_{H'}$ symmetry and (2) requiring a spectrum where the pion, $\pi^\pm$, is heavier than the $W^\pm$.
  \item The $\mu$-term in the scalar potential explicitly breaks the $\SU{(2)}_\text{A}$ axial symmetry. This gives a mass to the pion, which is a pseudo-Goldstone boson. This is analogous to the pion masses in chiral perturbation theory and the Higgs mass in composite Higgs models.
  \item Simultaneously requiring the pion to be heavy and the mediator light is a tuning of a dimensionful, renormalizable parameter. We take this to be $v_0^2$.
  \item The quartic terms set the mass of the radial modes with respect to the vevs. The validity of perturbation theory requires $\lambda, \lambda', \lambda'' \lesssim 4\pi$ and sets a maximum mass for these modes.
\end{itemize}
We sketch the spectrum in Fig.~\ref{fig:spectrum}.

\section{Symmetry Breaking}

A linear parameterization of the scalar fields is
\begin{align}
  H & =
  \begin{pmatrix}
    h_u \\
    h_d
  \end{pmatrix}
  &
  \Phi &=
  \frac 12
  \begin{pmatrix}
    \phi^3        & \sqrt{2} \phi^+ \\
    \sqrt{2} \phi^-  &  -\phi^3
  \end{pmatrix}
  &
  \phi^\pm \equiv \frac{\phi^1 \mp i\phi^2}{\sqrt{2}}
  \ .
\end{align}
We parameterize the vacuum expectation values of the fields by
\begin{align}
    \langle H \rangle & =
    \begin{pmatrix}
      0 \\
      v/\sqrt{2}
    \end{pmatrix}
    &
   \langle \Phi \rangle &=
    \frac 12
    \begin{pmatrix}
      f & \\
      & - f
    \end{pmatrix} = f T^3 \ .
    \label{eq:vevs}
\end{align}
These vevs break the global symmetries $\SU{(2)}_H \to \varnothing$ and $\SU{(2)}_\Phi \to \UU{(1)}$, respectively.

\subsection{Would-be Goldstones}

We parameterize the Goldstone fields as spacetime-dependent transformations of the vacuum by the broken generators~\cite{Callan:1969sn}:
\begin{align}
  H & =
    e^{i \frac{\varphi_H \cdot T}{v/2}}
    \langle H \rangle
  &
  \varphi_H \cdot T
  &=
  \sqrt{2}\varphi_H^+T^+
  +
  \sqrt{2}\varphi_H^- T^- + \varphi_H^0 T^3
  \label{eq:CCWZ:parameterization:H}
  \\
  \Phi & =
    e^{i \frac{\varphi_\Phi \cdot T}{f}}
    \,
    \langle \Phi \rangle
    \,
    e^{-i \frac{\varphi_\Phi \cdot T}{f}}
  &
  \varphi_\Phi \cdot T
  & =
  \sqrt{2}\varphi_\Phi^+ T^+
  +
  \sqrt{2}\varphi_\Phi^- T^-
    \ ,
    \label{eq:CCWZ:parameterization:phi}
\end{align}
with respect to the $\SU{(2)}_{H,\Phi}$ generators $T^\pm=T^1 \pm i T^2, T^3$. The radial modes are
  \begin{align}
    \left. H \right|_\text{radial} & =
    \frac{1}{\sqrt{2}}
    \begin{pmatrix}
      0 \\
      h
    \end{pmatrix}
    &
    \left. \Phi \right|_\text{radial} &=
    \frac 12
    \begin{pmatrix}
      \phi        & \\
      &  -\phi
    \end{pmatrix} \ .
  \end{align}

\subsection{Gauge Boson Masses}

The gauged $\SU{(2)}_V$ symmetry is the diagonal combination of $\SU{(2)}_H\times \SU{(2)}_\Phi$.
In our representation, the covariant derivatives on the scalar fields are
\begin{align}
D_\mu H
& = \partial_\mu H - igW^a_{\mu}T^a H
&
\mathcal{D}_\mu \Phi
& = \partial_\mu \Phi - igW^a_{\mu}\left[T^a,\Phi\right] \ .
\end{align}
If the fields acquire vevs \eqref{eq:vevs}, then the kinetic terms yield the following mass terms for the gauge bosons:
\begin{align}
  \mathcal L_\text{mass} & =
  m_W^2 W^+W^- + \frac 12 m_A^2 A^2
  &
  m_W^2 &= g^2 f^2 + \frac{g^2 v^2}{4}
  &
  m_A^2 &= \frac{g^2 v^2}{4} \ .
  \label{eq:gauge:boson:masses}
\end{align}
We identify the massive dark matter $W^\pm = (W^1 \mp iW^2)/\sqrt{2}$ and mediator (dark photon) $A = W^3$. The limit $v^2 \ll f^2$ yields a spectrum where the dark photon is much lighter than the dark matter.
The covariant derivatives with respect to the spin-1 mass eigenstates are
\begin{align}
D_{\mu}H
  & = \partial_{\mu}H-i\frac{g}{\sqrt{2}}\left(W^+_{\mu}T^+ +W^-_{\mu}T^-\right)H-i g A_{\mu}T^3 H
\\
\mathcal{D}_{\mu} \Phi
  &=  \partial_{\mu}\Phi-i \frac{g}{\sqrt{2}}
      \left(
        W^+_{\mu} [T^+,\Phi]+W^-_{\mu} [T^-,\Phi]
      \right)
      -i g A_{\mu} [T^3,\Phi] \ .
\end{align}

\subsection{Higgs Mechanism and Leftover Goldstones}
\label{section:sub:leftover:gold}

Let $\varphi_V$ be the linear combination of Goldstone bosons associated with $\SU{(2)}_V$.
Gauging the vector combination $\SU{(2)}_V$ promotes this global symmetry to a local symmetry. In unitary gauge one performs a local $\SU{(2)}_V$ transformation to remove $\varphi_V$ from the theory. It appears solely as the longitudinal polarization of the massive gauge bosons. We express $\varphi_V$ in terms of the $\varphi_{H,\Phi}$ by identifying this mixing in the kinetic terms:
\begin{align}
  |DH|^2 + \text{Tr}\,|\mathcal D\Phi|^2
  &
  \supset
  - g \left(\frac{v}{2} \partial\varphi_H^+ + f \partial \varphi_\Phi^+ \right) W^-  + \text{h.c.} - g \frac{v}{2} \partial\varphi_H^0 A
  \ .
  \label{eq:Goldstone:eating}
\end{align}
Only $\langle H \rangle$ breaks the $\UU{(1)}$ symmetry so that the photon $A$ eats the only neutral Goldstone. This is in contrast to the charged states for which there are two pairs of charged Goldstones and only one pair of charged gauge bosons. From \eqref{eq:Goldstone:eating} we identify the normalized $\SU{(2)}_V$ Goldstone $\varphi_V$ and the orthogonal state $\varphi_A$:
\begin{align}
  \varphi_V^\pm &=
  \frac{f \varphi_\Phi^\pm + (v/2)\varphi_H^\pm}{\sqrt{f^2 + (v/2)^2}}
  &
  \varphi_A^\pm &=
  \frac{f \varphi_H^\pm - (v/2)\varphi_\Phi^\pm}{\sqrt{f^2 + (v/2)^2}} \ . \label{eq:vectorandaxialstates}
\end{align}
Appendix~\ref{app:U1:example} presents an illustrative $\UU{(1)}$ example motivating these linear combinations. In unitary gauge, $\varphi_V^\pm$ only appears as the longitudinal mode of $W^\pm$. The `axial' combination $\varphi_A^\pm$ is an uneaten Goldstone boson that remains in the theory. We refer to these as pions and relabel them $\pi^\pm$ in anticipation of including explicit symmetry breaking terms to make them massive.

\subsection{Symmetry Breaking with $\lambda$, $\lambda'$ }

The simplest form of this model takes only the first two terms in \eqref{eq:V},
\begin{align}
\left.V\right|_{\lambda,\lambda'} & =
\phantom{+}
\frac{\lambda}{4!}
\left(
2 \tr \Phi^2 - f_0^2
\right)^2
+
\frac{\lambda'}{4!}
\left(
2|H|^2 - v_0^2
\right)^2 \ . \label{eq:V1}
\end{align}
These terms separately break the $\SU{(2)}_\Phi$ and $\SU{(2)}_H$ global symmetries. This, in turn, breaks the gauged vector combination of the two and gives mass to the gauge bosons. The vevs $f$ and $v$ are trivially related to the Lagrangian parameters $f_0$ and $v_0$,
\begin{align}
f &= f_0 &
v &= v_0 \ .
\label{eq:vev:0}
\end{align}
The radial modes modes $h$ and $\phi$ do not mix. Their masses are
\begin{align}
m_{\phi}^2 &= \frac{\lambda'}{3} f_0^2 &
m_h^2 &= \frac{\lambda}{3} v_0^2 \ .
\end{align}

\subsection{Symmetry Breaking with $\lambda$, $\lambda'$, $\mu$}
\label{eq:sym:br:lam:lamp:mu}

Introducing the $\mu$ term in the potential explicitly breaks $\SU{(2)}_{\Phi} \times \SU{(2)}_H \to \SU{(2)}_\text{V}$ and gives the $\pi^\pm$ a mass\footnote{In \acro{QCD}, the quark masses explicitly break chiral symmetry to give mass to the pions.} proportional to $\mu$:
\begin{align}
\left.V\right|_{\lambda,\lambda',\mu} & =
\phantom{+}
\frac{\lambda}{4!}
\left(
2 \tr \Phi^2 - f_0^2
\right)^2
+
\frac{\lambda'}{4!}
\left(
2|H|^2 - v_0^2
\right)^2
+
\mu H^{\dagger}\Phi H \ . \label{eq:V2}
\end{align}
This shifts the minimum of the potential from \eqref{eq:vev:0} to the  following condition:
\begin{align}
f^2&=f_0^2+\frac{3 \mu v^2}{2\lambda f}
&
v^2&=v_0^2+\frac{3 \mu f}{\lambda'}\ .
\label{eq:vevs2}
\end{align}
The $\mu$ term causes the $\Phi$ vev to shift the $H$ vev, and vice versa.

\subsubsection{Tuning for phenomenological hierarchy}

 Phenomenologically we require that the mediator is light and that the pions are heavier than the dark matter; this forces
\begin{align}
  g^2v^2 &\ll g^2f^2 \lesssim \mu f
 \ .
\end{align}
Assuming $g\lesssim \mathcal O(1)$, we see that the the vev $f^2$ is perturbed by a small amount relative to its $\mu=0$ value $f_0^2$. On the other hand, the hierarchy $f,\mu \gg v$ and perturbative limit $\lambda' < 4\pi$ imply that $v^2$ is shifted by a large amount relative to $v_0^2$. Without loss of generality, we assume $\mu>0$. We then require that $v_0^2$ is negative and tuned to give a small $v^2 \ll f^2$.

\subsubsection{Radial mode mixing}

The $\mu$-term induces mixing between the radial $\phi$ and $h$fields. Expanding \eqref{eq:V2} about the vacuum \eqref{eq:vevs2} yields a mass matrix $\mathcal{M}_H$,
\begin{align}
  \mathcal L
  & \supset
  \frac 12
  \begin{pmatrix}
    h & \phi
  \end{pmatrix}
  \mathcal{M}_H^2
  \begin{pmatrix}
    h \\ \phi
  \end{pmatrix}
&
\mathcal{M}_H^2
&=
\begin{pmatrix}
 \dfrac{\lambda' v^2}{3} &-\dfrac{\mu v}{2}\\
  -\dfrac{\mu v}{2} &  \dfrac{\lambda f^2}{3} +\dfrac{\mu v^2}{4 f}
\end{pmatrix}
\ .
\label{eq:MH}
\end{align}
The eigenvalues of the mass matrix are
\begin{align}
m_{1,2}^2
&= \frac{1}{2}\tr{\mathcal{M}_H^2}\left(1 \mp \sqrt{1-\frac{4 \Det{\mathcal{M}_H^2}}{\left(\tr{\mathcal{M}_H^2}\right)^2}}\right)\ . \label{eq:mpm}
\end{align}
We focus on the $v \ll f \sim \mu$ regime where the eigenvalues are positive%
\footnote{The minimum of the potential has positive squared masses. In \eqref{eq:mpm}, the possibility of a negative eigenvalue corresponds to the vev in \eqref{eq:vevs2} becoming a saddle point rather than a minimum. This occurs for large $\mu$ and is outside the regime of phenomenological interest for this study.}
 and have a large mass splitting. The light and heavy eigenvalues are
 \begin{align}
 m_1^2&=
 \frac{\lambda' v^2}{3}
 -\frac{3\mu^2 v^2}{4 \lambda f^2}
 +\mathcal{O}\left(\frac{v^4}{f^4}\right)
 &
 m_2^2&=
 \frac{\lambda f^2}{3}
 +\frac{\mu v^2}{4 f}
 +\frac{3 \mu^2 v^2}{4 \lambda f^2}+\mathcal{O}\left(\frac{v^4}{f^4}\right)
 \ . \label{eq:mpmorderv2}
 \end{align}
These correspond to light and heavy radial modes  that are a mixture of the $\phi$ and $h$ states:
\begin{align}
  \begin{pmatrix}
  H_1
  \\
  H_2
  \end{pmatrix}
  &=
  \mathbf{R}_\alpha
  \begin{pmatrix}
  h
  \\
  \phi
  \end{pmatrix}
  &
  \mathbf{R}_\alpha &=
  \begin{pmatrix}
  \phantom{+}\cos\alpha & \sin\alpha
  \\
  -\sin\alpha & \cos\alpha
  \end{pmatrix} \ .
  \label{eq:Hrotation}
\end{align}
The radial mode mixing angle is related to the model parameters by
\begin{align}
	\tan 2\alpha &=\frac{\mu v}{\lambda f^2/3 +\mu v^2 /4 f -\lambda' v^2/3}  \ . \label{eq:tan2alpha1}
\end{align}
\subsubsection{Goldstone mixing}

In addition to mixing the radial fields, the $\mu$ term mixes the charged Goldstones, $\varphi_{\Phi}^{\pm}$ and  $\varphi_{H}^{\pm}$. Expanding \eqref{eq:V2} yields a mass matrix
\begin{align}
  \mathcal L &\supset
  \begin{pmatrix}
    \varphi_{\Phi}^- & \varphi_H^-
  \end{pmatrix}
  \mathcal{M}_G^2
  \begin{pmatrix}
    \varphi_{\Phi}^+ \\
    \varphi_H^+
  \end{pmatrix}
&
\mathcal{M}_G^2
&=
\begin{pmatrix}
\phantom{+}\dfrac{\mu v^2}{4 f}& -\dfrac{\mu v}{2}\\
-\dfrac{\mu v}{2} & \phantom{+}\mu f
\end{pmatrix}
\ .
\label{eq:MG}
\end{align}
There is a massless mode because $\Det{\mathcal{M}_G^2}=0$. This corresponds to the massless Goldstones, $G^{\pm}$, eaten by charged gauge bosons $W^{\pm}$. The massive pions, $\pi^\pm$, have a mass-squared given by the trace:
\begin{align}
m_{G}^2&=0
&
m_{\pi}^2&=\mu f \left(1+\frac{v^2}{4 f^2}\right) \, . \label{eq:mpi}
\end{align}
The mass eigenstates are related to the would-be Goldstones, $\varphi_{\Phi}^\pm$ and $\varphi_{H}^\pm$, by a rotation
\begin{align}
  \begin{pmatrix}
  G^{\pm}
  \\
  \pi^{\pm}
  \end{pmatrix}
  &=
  \mathbf{R}_{\beta}
  \begin{pmatrix}
  \varphi_{\Phi}^{\pm}
  \\
  \varphi_H^{\pm}
  \end{pmatrix}
  &
  \mathbf{R}_{\beta}=
  \begin{pmatrix}
  \phantom{+}\cos\beta & \sin\beta
  \\
  -\sin\beta & \cos\beta
  \end{pmatrix} \ .
  \label{eq:Grotation}
\end{align}
The Goldstone mixing angle, $\beta$, satisfies
\begin{align}
\tan\beta & = \frac{v}{2 f}
&
\sin\beta & = \frac{v/2}{\sqrt{f^2+v^2/4}}
&
\cos\beta & =\frac{f}{\sqrt{f^2+v^2/4}} \ ,
\label{eq:tanbeta}
\end{align}
where we assume $0 \leq \beta \leq \pi/2$. Observe that in the absence of a $\mu$ term, the mass eigenstates in \eqref{eq:Grotation} are identical to those defined by \eqref{eq:vectorandaxialstates}. This shows that the gauging of the vector combination $\SU{(2)}_V$ fixes a basis of eaten Goldstones, $G^\pm$, and their orthogonal states, $\pi^\pm$; see Appendix~\ref{app:U1:example}. The latter non-linearly realize $\SU{(2)}_A$ and pick up an explicit mass when we introduce the $\mu$ term.

\subsection{Symmetry Breaking with $\lambda$, $\lambda'$, $\lambda''$, $\mu$ }

The most general renormalizable potential \eqref{eq:V} includes a mixed quartic term $\lambda'' |H|^2\text{Tr}\Phi^2$. This term shifts the vevs and affects the radial mode mixing but does not induce any further Goldstone interactions since it is manifestly $\SU{(2)}_H\times \SU{(2)}_\Phi \times \UU{(1)}_H$ invariant.
The vevs in this scenario are shifted from \eqref{eq:vevs2}:
\begin{align}
	f^2&=f_0^2+\frac{3 v^2}{\lambda}\left(\frac{\mu}{2 f}-\lambda''\right)
	&
	v^2&=v_0^2+\frac{3 f^2}{\lambda'}\left(\frac{\mu}{f}-\lambda''\right)\ . \label{eq:vevs3}
\end{align}
The $\lambda''$ term introduces additional interactions and mixing between the radial modes. The radial field mass matrix is
\begin{align}
	\mathcal{M}_H^2 &=
	\begin{pmatrix}
	\dfrac{\lambda' v^2}{3}&\lambda'' v f-\dfrac{\mu v}{2}\\
	\lambda'' v f-\dfrac{\mu v}{2} & \dfrac{\lambda f^2}{3} +\dfrac{\mu v^2}{4 f}
	\end{pmatrix}. \label{eq:MH2}
\end{align}
The eigenvalues of \eqref{eq:MH2} are given by \eqref{eq:mpm} and yield
\begin{align}
m_1^2&= \frac{\lambda' v^2}{3}-\frac{3 v^2}{4 \lambda f^2}\left(\mu-2 f \lambda''\right)^2
+\mathcal{O}\left(\frac{v^4}{f^4}\right)
\label{eq:mpmorderv2_2:1}
\\
m_2^2 &= \frac{\lambda f^2}{3}+\frac{\mu v^2}{4 f}+\frac{3 v^2}{4 \lambda f^2}\left(\mu-2 f \lambda''\right)^2
+\mathcal{O}\left(\frac{v^4}{f^4}\right)
\ . \label{eq:mpmorderv2_2}
\end{align}
The rotation \eqref{eq:Hrotation} that diagonalizes \eqref{eq:MH2} is modified from \eqref{eq:tan2alpha1} to
\begin{align}
	\tan 2\alpha =\frac{\mu v-2\lambda'' v f}{\lambda f^2/3 +\mu v^2 /4 f -\lambda' v^2/3}\ . \label{eq:tan2alpha2}
\end{align}
For the remainder of this manuscript we set $\lambda''=0$ since its primary phenomenological effects may be understood as a shift on $\mu$.

\subsection{Qualitative Behavior}

The parameters of interest realize the spectrum in Fig.~\ref{fig:spectrum}. The qualitative behavior of the theory is the limit where the longitundinal $W$ modes are predominantly the triplet Goldstones and
\begin{itemize}
  \item $\alpha = 0$\;: the light Higgs is predominantly the doublet neutral Goldstone,
  \item $\beta = 0$\;: the charged pions are predominantly the doublet charged Goldstones.
\end{itemize}

\subsection{Vacuum Stability}
\label{sec:vac:stability}

The stability of the vacuum requires $\det\mathcal{M}_H^2>0$. This implies the inequality
\begin{align}
	\frac{\lambda'}{9}
  \left(
    \lambda+\frac{3 \mu}{f}\frac{v^2}{4 f^2}
    \right)
  & >
  \left(
    \lambda''-\frac{\mu}{2 f}
    \right)^2
    \label{eq:stability}.
\end{align}
The $v\ll f$ limit implies a maximal value of the trilinear mass scale, $\mu_{\text{max}}$. For $\mu$ larger than this the $v \ll f$ critical point is a saddle point; there still exists a stable minimum however it does not realize the limit $v \ll f$.

\section{Feynman Rules for Light States}

We summarize the dark sector Feynman rules for the dark matter and the low-mass states. The dark matter and dark photon have interactions analogous to the Standard Model $W$ and $Z$ bosons and are thus given by
\begin{align}
  \vcenter{
		\hbox{\includegraphics[width=.2\textwidth]{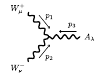}}
		}
	&=
  -i g
  \left[
    g^{\mu \nu}\left(p_1 -p_2\right)^{\lambda}+g^{\nu \lambda}\left(p_2-p_3\right)^{\mu}+g^{\lambda \mu}\left(p_3-p_1\right)^{\nu}
    \right]
  \label{eq:darkphotonDMrule}
  \\
	\vcenter{
		\hbox{\includegraphics[width=.2\textwidth]{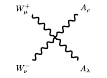}}
		}
	&=
  -i g^2
  \left[
    2 g^{\mu \nu}g^{\lambda\rho}-g^{\mu \lambda}g^{\nu \rho}-g^{\mu \rho}g^{\nu \lambda}
    \right]\ .
    \label{eq:darkphotonDMcontactrule}
\end{align}
The Feynman rules for the dark matter interactions with the light radial mode are
\begin{align}
		\vcenter{
  		\hbox{\includegraphics[width=.2\textwidth]{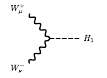}}
  		}
		&=
    i g m_W g^{\mu \nu}
    \left(
      \cos\alpha \sin\beta +2\sin\alpha \cos\beta
      \right)
    \label{eq:H1DMrule}
    \\
		\vcenter{
  		\hbox{\includegraphics[width=.2\textwidth]{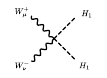}}
  		}
		&=
    \frac{i g^2}{4}
    \left(
      5-3\cos 2\alpha
      \right) g^{\mu \nu}\ .
      \label{eq:H1DMcontactrule}
\end{align}
With respect to the general renormalizable spin-1 dark matter Lagrangian parameterization in \cite{Dent:2015zpa,Catena:2019hzw,Catena:2018uae}, these rules correspond to
$b_5 = g$ and $b_6 = ig$, with other identifications straightforward.

\section{Relic Abundance and Annihilation}
\label{sec:relic}

\begin{figure}[t]
	\begin{center}
	\includegraphics[width=\textwidth]{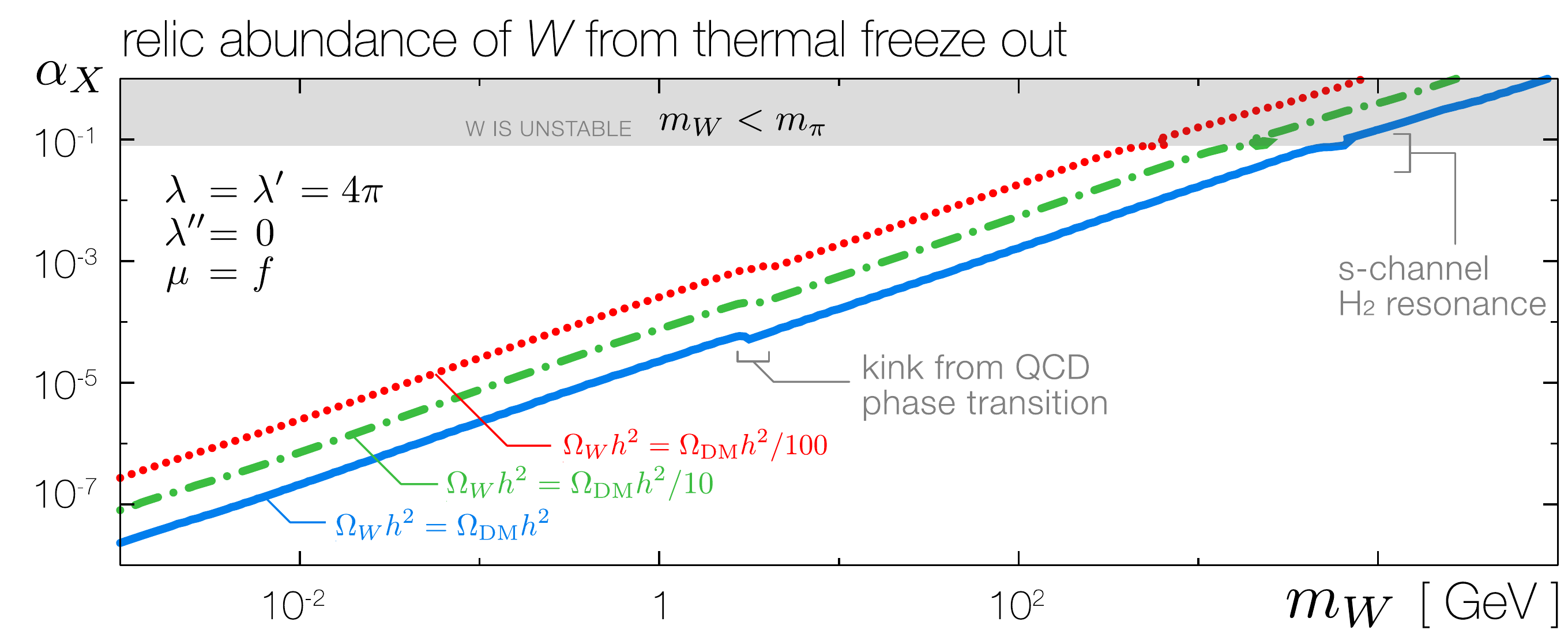}
	\end{center}
\caption{The $W$ relic abundance from thermal freeze out as a function of the $W$ mass and fine structure constant. We plot lines where the $W$ saturates the entire dark matter density (solid/\textsf{\color{blue}blue}) or only a 10\% (dash-dotted/\textsf{\color{ForestGreen}green}) or 1\% fraction (dotted/\textsf{\color{red}red}). We take $\lambda=\lambda'=4\pi$, $\lambda''=0$ and $\mu=f$. The shaded region is excluded in order to prevent dark matter decay.}
\label{fig:relicdensity}
\end{figure}

\begin{figure}
  \begin{center}
  \includegraphics[width=.22\textwidth]{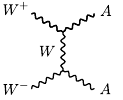}
  \;
  \includegraphics[width=.22\textwidth]{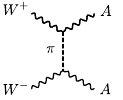}
  \;
  \includegraphics[width=.22\textwidth]{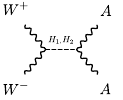}
  \;
  \includegraphics[width=.22\textwidth]{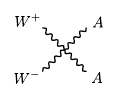}
\end{center}
\caption{Diagrams contributing to $W^+ W^- \to A A$ annihilation. Not shown: crossed ($u$-channel) diagrams and annihilation to scalars.}
\label{fig:WWAAdiagrams}
\end{figure}

Dark matter annihilation is dominated by $s$-wave processes that persist in the zero-relative-velocity limit. We sketch the primary $W^+W^-\to A A$ diagrams in Fig.~\ref{fig:WWAAdiagrams}. The largest $s$-wave channels are $W^+W^-$ annihilating to $AA$ and $H_1H_1$. The $AH_1$ final state vanishes in the $m_A/m_W \to 0$ limit. Note that we asssume that the entropy produced in dark matter annihilation is eventually dumped into the visible sector through the portal interactions in Section~\ref{sec:portal}. The relevant annihilation cross sections are:
\begin{align}
	\sigma v_{AA}
  =&
	\frac{\pi \alpha_X^2 }{36 m_{W}^2}
  \left\{ 152+\frac{4 m_{W}^4}{\left(m_{W}^2+m_{\pi}^2\right)^2}+\frac{3 \left(-4 m_{W}^2+m_2^2+2 m_{\pi}^2\right)^2}{4 \left(m_2^2-4 m_{W}^2\right)^2}\right. \nonumber
  \\
	&\qquad\qquad
  \left.+\frac{2 m_{W}^2 \left(-4 m_{W}^2+m_2^2+2 m_{\pi}^2\right)}{\left(m_{W}^2+m_{\pi}^2\right) \left(4 m_{W}^2-m_2^2\right)}
  \right\}\label{eq:sigmaAA}
  \\
  \sigma v_{H_1 H_1}
  =&\frac{\pi  \alpha_X ^2}{144 m_{W}^2}
  \left\{3+\frac{12 m_{\pi}^4}{\left(m_2^2-4 m_{W}^2\right){}^2}-\frac{16 m_{\pi}^2 m_{W}^2}{\left(m_2^2-4 m_{W}^2\right) \left(m_{W}^2+m_{\pi}^2\right)}+\frac{16 m_{W}^4}{\left(m_{W}^2+m_{\pi}^2\right)^2}
  \right\} \ .
  \label{eq:sigmaH1H1}
\end{align}
We define the dark fine structure constant
\begin{align}
  \alpha_X &= \frac{g^2}{4\pi} \ .
\end{align}
In the decoupling limit where both $m_{\pi}$ and $m_2 \to \infty$, \eqref{eq:sigmaAA} matches the calculation for a spin-1 dark matter particle annihilating into massless dark photons in Ref.~\cite{Baek:2013dwa}. For completeness, we list the $s$-wave annihilation cross sections going into final states with the heavy Higgs, $H_2$, though these are typically kinematically suppressed. The relevant final states are $H_2H_2$ and $AH_2$; these are only allowed when $m_{W}>m_2$ and $m_{W}>m_2/2$ respectively. The $H_1H_2$ mode vanishes in the $m_A/m_W \to 0$ limit.
\begin{align}
\sigma v_{H_2 H_2}=&\frac{2\pi \alpha_X^2}{9 m_{W}^2}\sqrt{1-\frac{m_2^2}{m_{W}^2}}\frac{864 m_{W}^8+31 m_2^8-248 m_2^6 m_{W}^2+820 m_2^4 m_{W}^4-1296 m_2^2 m_{W}^6}{\left(-6 m_2^2 m_{W}^2+8 m_{W}^4+m_2^4\right)^2}\label{eq:sigmaH2H2}\\
\sigma v_{A H_2} =&\frac{8 \pi  \alpha_X ^2 }{9 m_{W}^4}\left(4 m_{W}^2-m_2^2\right) \label{eq:sigmaAH2} \ .
\end{align}

We assume that the relic abundance is set by non-relativistic freeze out. The freeze-out temperature and final relic abundance is~\cite{Kolb:1990vq}
\begin{align}
	x_f
  =&
  \ln \left[0.038\sqrt{\frac{g}{g_*}}M_{Pl} \, m_{W} \langle\sigma v\rangle \right]
  -
  \frac{1}{2}\ln^2 \left[0.038\sqrt{\frac{g}{g_*}} M_{Pl}\, m_{W}\langle\sigma v\rangle\right]
  \label{eq:xf}
  \\
	\Omega h^2 =& 2\times1.07 \times 10^9 \frac{x_f~\text{GeV}^{-1}}{\sqrt{g_*(x_f)}M_{Pl}\langle\sigma v\rangle} \label{eq:relicdensity} \ ,
\end{align}
where we include an explicit factor of two in \eqref{eq:relicdensity} to account for a given dark matter particle, $W^\pm$, only being able to annihilate with its anti-particle, $W^\mp$. This is compared to $\Omega h^2 = 0.12$~\cite{Tanabashi:2018oca, Steigman:2012nb}. Fig.~\ref{fig:relicdensity} shows the coupling $\alpha_X = g^2/4\pi$ that reproduces the observed dark matter relic abundance assuming thermal freeze out for benchmark parameters.
For fermionic dark matter annihilating into much lighter dark photons, a numerical estimate for the target fine structure constant is
$\alpha_{X,\text{fermion}}^\text{th} \cong 0.035~(m_X/\text{TeV})$ (see e.g.~\cite{Liu:2014cma}). Comparing the $m_A \to 0$ fermionic $XX\to AA$ cross section to \eqref{eq:sigmaAA}:
\begin{align}
  \langle \sigma_{XX\to AA} v \rangle^\text{fermion} &\approx \frac{\pi\alpha_X^2}{m_X^2}
  &
  \langle \sigma_{WW\to AA} v \rangle &\approx \frac{38}{9}  \frac{\pi\alpha_X^2}{m_X^2} \ .
\end{align}
We thus estimate the target $\alpha_X$ in our model by rescaling the fermionic target by $(38/9)^{-1/2} \approx 0.5$:
\begin{align}
  \alpha_X^\text{th} \cong 0.017~\left(\frac{m_W}{\text{TeV}}\right) \ . \label{eq:alphath}
\end{align}
This estimate ignores the contributions from $H_1$ final states or possible $H_2$ resonances (see Fig.~\ref{fig:alphavsmpi}).

Implicit in our assumption is that the dark photon, $A$, is sufficiently in equilibrium with the Standard Model. We thus assume
\begin{equation}
	\Gamma_A \geq H(x_f\cong 20)
\end{equation}
where $\Gamma_A$ is the dark photon decay width and $H(x_f)$ is the Hubble rate evaluated at freeze-out. This places a lower bound on the kinetic mixing with the visible sector, $\varepsilon$ in (\ref{eq:Lkineticmixing}) \cite{Pospelov:2007mp}:
\begin{equation}
	\varepsilon^2 \left(\frac{m_A}{10~\text{MeV}}\right)\gtrsim 10^{-11}\left(\frac{m_W}{50~\text{GeV}}\right)^2 \ .
\end{equation}
This is not strictly necessary. One simple direction is to assume a dark sector with a completely different initial temperature at reheating~\cite{Feng:2008mu}. More generally, the full `phase space' of thermal histories for dark sectors with mediators is an exciting direction that only recently been studied~\cite{Chu:2011be,Blennow:2013jba,Bernal:2015ova,Krnjaic:2017tio,Evans:2017kti, Dvorkin:2019zdi}. 
Alternatively, one may pursue models where \acro{UV} dynamics produce asymmetric dark matter within our scenario~\cite{Kaplan:2009ag,Turner:1987pp}.
These possibilities are beyond the scope of the present work. Here we focus on the simple benchmark scenario where the $W$ abundance is produced through standard thermal freeze out by annihilation into mediators. Explorations of the alternative scenarios are especially interesting and we leave them for future work.

\section{Relating Dark Matter and Dark Pion Masses}

\begin{figure}
	\begin{center}
    \includegraphics[width=.45\textwidth]{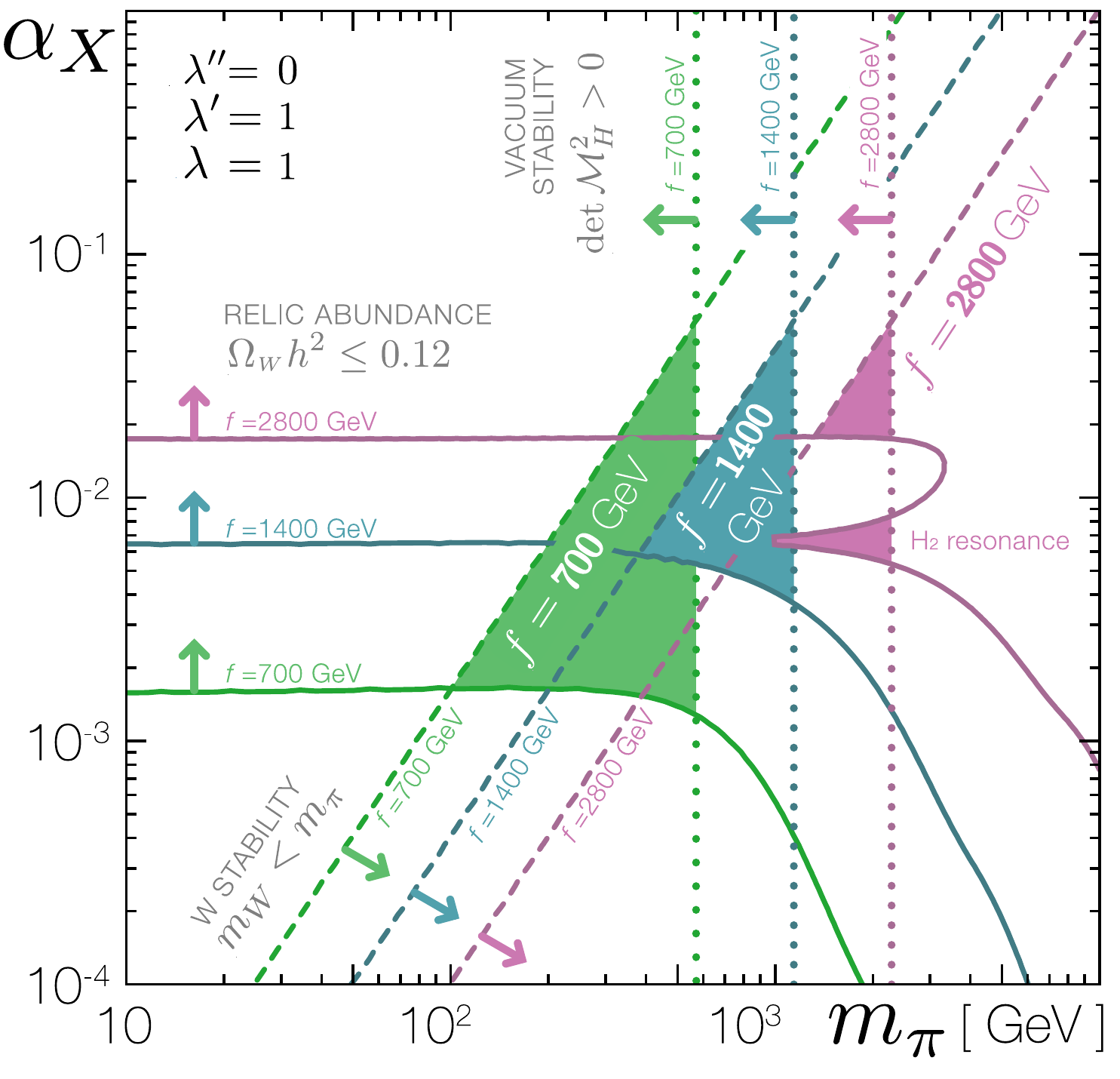}
    \qquad
    \includegraphics[width=.45\textwidth]{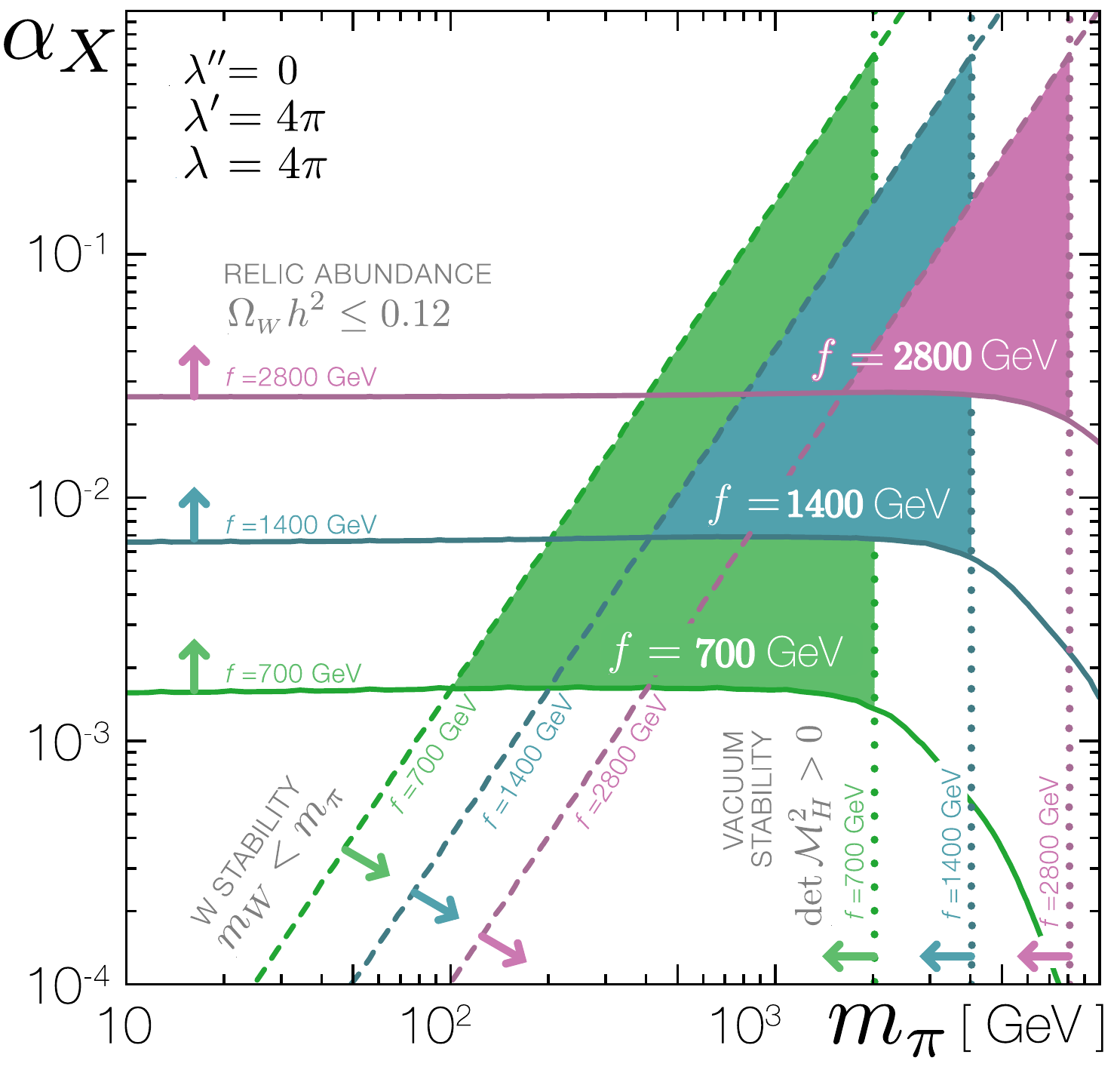}
	\end{center}
	\caption{Shaded regions correspond to values of the dark fine structure constant $\alpha_X = g^2/4\pi$ and the pion mass $m_\pi \approx \mu f$ that (\emph{i}) do not overclose the universe [\textsf{solid lines}], (\emph{ii}) have a stable $W$ [\textsf{dashed lines}], and (\emph{iii}) have a stable vacuum [\textsf{dotted lines}]. Colors correspond to choices of $f=700$ GeV (\textsf{\color{ForestGreen}green}, lower-left), $f=1400$ GeV (\textsf{\color{NavyBlue}teal}, middle), $f=2800$ GeV (\textsf{\color{Plum}magenta}, upper-right). We take $\lambda''=0$ for simplicity. \textsc{Left}: $\lambda = \lambda' = 1$. \textsc{Right}: $\lambda=\lambda'=4\pi$.
	\label{fig:alphavsmpi}
  }
\end{figure}

In the $v\ll f$ limit of phenomenological interest, the properties of the dark matter $W$ and the pions $\pi^\pm$ are related to one another.
Fig.~\ref{fig:alphavsmpi} shows the allowed region for $m_{\pi} \approx \mu f$ and $\alpha_X=g^2/4\pi$ for a sample of vevs, $f$. The dark matter mass is $m_W \approx \sqrt{4\pi \alpha_X}f$. The triangular regions are bounded by requiring
\begin{enumerate}
  \item a relic abundance less than or equal to the total dark matter abundance (Section~\ref{sec:relic}),
  \item the $W^\pm$ is the lightest charged particle in the dark sector (Section~\ref{eq:sym:br:lam:lamp:mu}), and
  \item the tree-level stability of the vacuum (Section~\ref{sec:vac:stability}).

\end{enumerate}
Observe the following features in Fig.~\ref{fig:alphavsmpi}:
\begin{itemize}
  \item As the symmetry breaking scale $f$ is increased, the $W$ mass increases so that the required coupling to saturate the dark matter relic abundance increases, the minimum pion mass increases to maintain the particle spectrum increases, and the bound on the stability of the scalar vevs \eqref{eq:stability} shifts to larger $\mu$ and hence larger $m_\pi$. Note that the stability bound is modified if $\lambda''> 0$.
  \item In the left-hand plot ($\lambda=\lambda'=1$), for $f=$~2800~GeV, the relic abundance bound exhibits a resonance in the annihilation diagram with an $s$-channel heavy Higgs, $H_2$. This is a useful reminder that the dynamics $H_2$ may be dominant in certain annihilation channels even though it will not affect the other observational probes discussed in this manuscript.
  \item Comparing the left-hand ($\lambda = \lambda' = 1$) and right-hand left-hand ($\lambda = \lambda' = 4\pi$) plots, the $W$ stability lines are unchanged. The other two bounds shift according to the $\lambda^{(')}$-dependence of the radial mode masses, (\ref{eq:mpmorderv2_2:1}--\ref{eq:mpmorderv2_2}), and the dependence of the annihilation rate and vacuum stability condition on these masses.
  \item In the unstable $W$ region ($m_W > m_\pi$), the relic abundance curves are flat and independent of the pion mass.  This corresponds to the leading terms in (\ref{eq:sigmaH1H1}--\ref{eq:sigmaAA}) that are $m_\pi$-independent. Note that the left- and right-handed plots differ in this flat region since $m_1$ depends on $\lambda'$ via \eqref{eq:mpmorderv2_2:1} so that increasing $\lambda'$ decreases the phase space for $W^+W^-\to H_1H_1$, resulting a slightly larger $\alpha_X$ required to annihilate enough $W$s.
  \item In the stable $W$ region ($m_W < m_\pi$) the $m_\pi$ dependence of the annihiliation cross sections (\ref{eq:sigmaAA}--\ref{eq:sigmaH1H1}) manifests itself. In this regime, the left- and right-hand plots differ in $m_\pi$ dependence because of the $\lambda$-dependence through the heavy Higgs mass, \eqref{eq:mpmorderv2_2}.
\end{itemize}

\section{Self-Interacting Dark Matter}

The dark sector furnished by our framework automatically realizes the self-interacting dark matter (\acro{SIDM}) paradigm. Ref.~\cite{Carlson:1992fn} first proposed that dark matter may exist in a separate sector with self-interactions. Refs.~\cite{Spergel:1999mh,Dave:2000ar} identified that the self-interactions may affect the density profiles of dark matter halos and thus allow observational tests of the dark matter self-interaction cross section. More recently, the seminal work in Refs.~\cite{Feng:2009mn,Feng:2009hw,Buckley:2009in,Tulin:2013teo} connected particle physics models of dark sectors (dark matter with low-mass mediators) to observed small scale structure anomalies tied to the dark matter density profiles of dwarf galaxies. We refer to Ref.~\cite{Tulin:2017ara} for a review.

The exchange of dark photons, $A$, generates a long range, velocity-dependent, self-interaction between the $W^{\pm}$ dark matter particles. At low energies, these self-interactions produce a Yukawa potential,
\begin{align}
  V(r)
  &=
  \pm \frac{\alpha_X}{r}e^{-m_A r}\ .
  \label{eq:Yukawapotential}
\end{align}
Since the force mediator is a vector boson, particle--antiparticle interactions produce an attractive potential while particle-particle interactions produce a repulsive potential. The self-interaction potential also receives contributions from the exchange of the radial modes, $H_1$ and $H_2$, that are purely attractive. We assume that both of these contributions are negligible:
\begin{align}
  \vcenter{
    \hbox{\includegraphics[width=.2\textwidth]{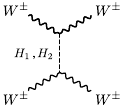}}
    }
    &\quad
    \ll
    \quad
  \vcenter{
    \hbox{\includegraphics[width=.2\textwidth]{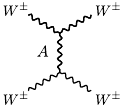}}
    } \ .
\end{align}
The heavy Higgs, $H_2$, is typically much heavier so that the Yukawa suppression causes the force to be short ranged. The light Higgs, $H_1$, is assumed to be heavier than the dark photon but may have a mass of the same order of magnitude. In this case, we note that the $H_1$ exchange diagram is suppressed by a factor of $(m_A/m_W)^2$ relative to $A$ exchange. This suppression is clear in the Feynman rule \eqref{eq:H1DMrule} where we note that $\sin\alpha \sim \sin\beta \sim v/f$ from \eqref{eq:tan2alpha1} and \eqref{eq:tanbeta}.

The long-ranged potential \eqref{eq:Yukawapotential} is the same as that generated by more conventional spin-1/2 or spin-0 models of self-interacting dark matter so that the phenomenology is qualitatively similar. A benchmark model in the conventional scenario is a 15~GeV dark matter with a 17~MeV mediator~\cite{Kaplinghat:2015aga}. The target cross section for this scenario is $\sigma \sim 1~\text{cm}^2~(m_X/g)$ for dwarf-scale velocities; this flattens the dark matter density in galactic cores~\cite{2012MNRAS.423.3740V, Rocha:2012jg, 2013MNRAS.431L..20Z}. This potential manifests a velocity dependence depending on the value of the transfer momentum compared to the mass of the mediator. This velocity-dependence suppresses the effect of self-interactions for large systems such as colliding galaxy clusters, where there is little evidence for self-interactions.

We compare the effects of the long-ranged dark matter self-interaction in our model with respect to the standard \acro{SIDM} benchmark.
One difference in our scenario is that we assume that dark matter is symmetric: it is composed of equal parts of $W^+$ and $W^-$. Cosmological constraints on the matter power spectrum constrain the early-universe annihilation of dark matter in the standard self-interacting dark matter scenario~\cite{Huo:2017vef}\footnote{We leave an exploration of these constraints for future work; in this manuscript we focus on the presentation of the core model with thermal freeze out.}. As such, the most viable \acro{SIDM} models typically assume that the dark matter is asymmetric to avoid these bounds. This assumption, in turn, implies that dark matter self-scatters are purely repulsive and avoid resonances.

\begin{figure}[t]
	\begin{center}
		\includegraphics[width=\textwidth]{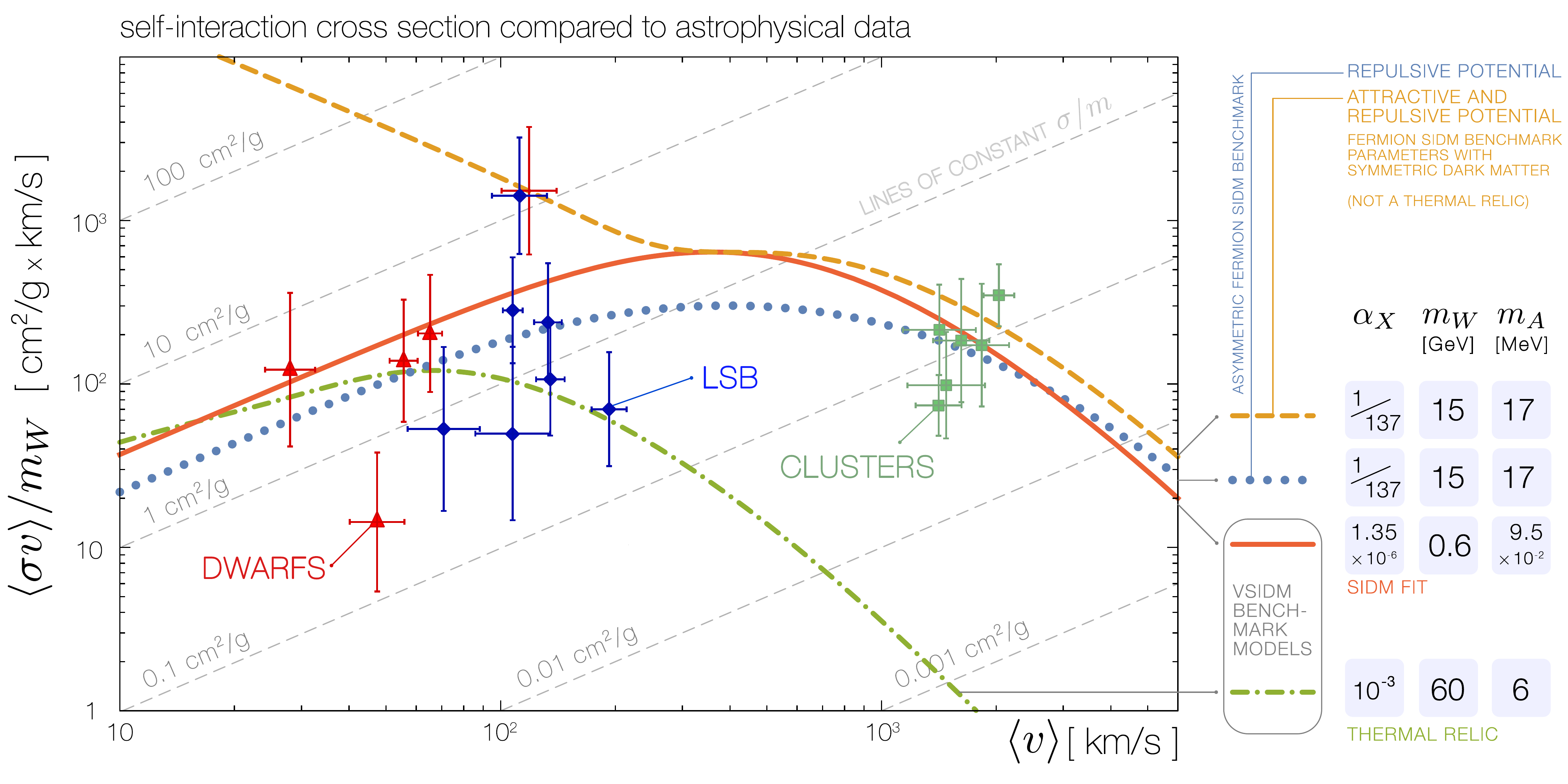}
	\end{center}
	\caption{Self-interaction cross section as a function of average velocity in our model compared to inferred cross sections for a set of dwarf galaxies, low surface brightness (\acro{LSB}) spiral galaxies, and galaxy clusters from Ref.~\cite{Kaplinghat:2015aga}.
  %
  %
  \textbf{Benchmark models}: the solid/\textsf{\color{red}red} curve are a fit to the inferred cross sections from astrophysical data. The $W^{\pm}$ are required to have a sub-GeV mass in order to agree with the cluster scale observations~\cite{Huo:2017vef}. The dash-dotted/\textsf{\color{ForestGreen}green} curve corresponds to a fit to the low-velocity data points while maintaining a GeV scale mass. Both benchmark models are subject to the requirement that $\alpha_X$ is large enough for the $W^\pm$ to saturate the dark matter relic abundance; see Fig.~\ref{fig:relicdensity}.
\textbf{Illustrative unphysical examples}: the dotted/\textsf{\color{blue}blue} line assumes a \emph{purely repulsive potential} and reproduces the best fit curve from Ref.~\cite{Kaplinghat:2015aga} using the same model parameters. The dashed/\textsf{\color{orange}yellow} line corresponds to the same model parameters but including both attractive and repulsive potentials.}
	\label{fig:SIDMplot}
\end{figure}

In this manuscript, we focus on benchmark models of symmetric vector self-interacting dark matter with both attractive and repulsive interactions. We plot the velocity-dependence of the self-interaction cross section, $\langle \sigma v\rangle$, in Fig.~\ref{fig:SIDMplot}. This reproduces the data from Fig.~1 of Ref.~\cite{Kaplinghat:2015aga} overlaid with curves based on our model. The methodology for producing these cross section curves is based on Ref.~\cite{Tulin:2013teo}; we present a self-contained summary in Appendix~\ref{app:SIDM:method}. The two benchmark parameters are:
\begin{enumerate}
  \item \textbf{The solid/\textsf{\color{red}red} curve is an  estimated fit to the inferred self-interaction cross sections for the astrophysical systems}. The dark matter mass $m_W$ is chosen to be 60~MeV in order to satisfy constraints on cluster scales~\cite{Kaplinghat:2015aga,Huo:2017vef}. The coupling is then fixed by \eqref{eq:alphath}. This model fits the \acro{SIDM} targets and is able to explain the dark matter abundance from thermal freeze out.
  \item \textbf{The dash-dotted/\textsf{\color{ForestGreen}green} curve is a model constrained to the observed dark matter relic density for a weak scale mass}. The coupling and dark matter mass are three orders of magnitude stronger than that of the \acro{SIDM} fit. This case is a reasonable fit for the inferred \acro{SIDM} cross sections from dwarfs and low surface brightness spiral galaxies, but falls orders of magnitude short of the inferred cross section from galaxy cluster profiles.
\end{enumerate}
In addition to these two benchmark models, we present two illustrative curves to highlight important physics:
\begin{enumerate}
  \setcounter{enumi}{2}
  \item \textbf{The dotted/\textsf{\color{blue}blue} curve shows a fit assuming only a repulsive potential}. This model reproduces the best fit curve from Ref.~\cite{Kaplinghat:2015aga} with the same model parameters. The spin of the dark matter candidate makes no appreciable difference since the long-range self-interaction potential is identical. However, since we consider symmetric dark matter, the assumption of a purely repulsive potential is unphysical.
  \item \textbf{The dashed/\textsf{\color{orange}yellow} curve shows the same model parameters as the dotted/\textsf{\color{blue}blue} curve, but with both attractive and repulsive interactions}. If one simply turns on the attractive contribution to the dotted/\textsf{\color{blue}blue} curve, one can see the effect of resonances. The cross section increases rapidly for low velocities and is a poor fit for the data. Comparing to the dash-dotted/\textsf{\color{ForestGreen}green} curve, we see that a modest shift in the model parameters is sufficient to move off of the resonance.
\end{enumerate}

\section{Portal Interactions}
\label{sec:portal}

We discuss renormalizable portal interactions between the dark sector and the Standard Model. In this context, our convention of naming particles by their Standard Model analogs can be ambiguous. For consistency and clarity, we write the visible sector fields in script font; see Table~\ref{tab:notation}.
\begin{table}\centering
  \begin{tabular}{lllllll}
    \toprule
     & \phantom{abc}
     & \multicolumn{2}{l}{\small\textsc{Dark Sector}}
     & \phantom{abc}
     & \multicolumn{2}{l}{\small\textsc{Standard Model}}
     \\
     \cmidrule{3-4}
     \cmidrule{6-7}
     Description
     &
     & Symbol & Name
     &
     & Symbol & Name
     \\ \midrule
     Charged \SU{(2)} gauge boson
     &
     & $W^\pm$ & dark matter
     &
     & $\mathcal{W}^\pm$ & $W$-boson
     \\
     Light neutral gauge boson
     &
     & $A$ & dark photon
     &
     & $\mathcal{A}$ & photon
     \\
     Heavy neutral gauge boson
     &
     &  &
     &
     & $\mathcal{Z}$ & $Z$-boson
     \\
     Light radial (Higgs) mode
     &
     & $H_1$  & dark Higgs
     &
     & $\mathfrak{h}$ & Higgs boson
     \\
     Heavy radial (Higgs) mode
     &
     & $H_2$  & heavy Higgs
     &
     &  &
     \\
     Charged pseudo-Goldstone
     &
     & $\pi^\pm$  & dark pion
     &
     & $\Pi^\pm$ & charged pion
     \\
     Neutral pseudo-Goldstone
     &
     &  &
     &
     & $\Pi^0$ & neutral pion
     \\
    \bottomrule
  \end{tabular}
  \caption{
  Conventions for dark sector and visible sector mass eigenstates.
  \label{tab:notation}
  }
\end{table}

The dark sector doublet $H$ and triplet $\Phi$ may have renormalizable interactions with the Standard Model Higgs $\mathcal H$ through mixed quartics:
\begin{align}
  \mathcal L \supset
  \lambda_{H\mathcal H} |H|^2|\mathcal H|^2  +
  \lambda_{\Phi\mathcal H} \left(\text{Tr}\,\Phi^2\right)|\mathcal H|^2 \ .
\end{align}
This leads to Higgs portal interactions of the type described in~\cite{Baek:2013dwa}\footnote{Our scenario differs slightly in that the low-mass dark Higgs is a mixture that is mostly composed of the radial mode of a doublet rather than a triplet.}.  In this manuscript we instead focus on the limit where the Higgs portal interactions are negligible\footnote{In the limit where this interaction is taken to be zero, the $H_1$ to $AA$ is at loop level.} compared to the dimension-5 operator,
\begin{align}
  \mathcal L \supset
  \frac{2}{\Lambda}
  \left(\Phi^a F^a_{\mu\nu} \right) \mathcal{B}^{\mu\nu} \ ,
  \label{eq:kin:mix:dim:5}
\end{align}
where $\mathcal{B}^{\mu\nu}$ is the Standard Model hypercharge field strength and $\Lambda$ encodes the combination of couplings and a \acro{UV} scale at which this term is generated by additional dynamics, for example heavy particles running in a loop. The vev $\langle\Phi\rangle = fT^a$ induces a kinetic mixing between the dark photon and the visible photon~\cite{Chen:2009dm}:
\begin{align}
  	\mathcal{L} \supset \frac{\varepsilon}{2 \cos\theta_W}F_{\mu \nu}\mathcal B^{\mu \nu}
    &\rightarrow  \frac{\varepsilon}{2} F_{\mu \nu}\mathcal F^{\mu \nu} \ ,
     \label{eq:Lkineticmixing}
\end{align}
where $\mathcal F_{\mu \nu}$ is the visible sector photon field strength. We have omitted a mixing term with the $Z$-boson field strength which exists in principle but is negligible in the limit where the dark photon is much lighter than the electroweak scale, $m_A \ll m_Z$; we refer to \cite{Izaguirre:2015eya} or the appendix of \cite{Feng:2016ijc} for a detailed derivation. We ignore the limit where additional symmetry breaking leads to dark $Z$ phenomenology~\cite{Davoudiasl:2012ag}.
By focusing on this kinetic mixing scenario~\cite{Holdom:1985ag,Galison:1983pa,Kobzarev:1966qya}, we study the hitherto unexplored case of vector dark matter interacting through a low-mass vector mediator.

The dark photons in our scenario are identical to the standard set up in how they interact with visible sector fields and, thus, how experiments may search for them~\cite{Battaglieri:2017aum,Alexander:2016aln}. The effective Feynman rule to fermions $f$ with electric charge $Q_f$, for example, is
\begin{align}
  \vcenter{
    \hbox{\includegraphics[width=.2\textwidth]{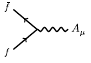}}
    }
  &=
  i \varepsilon e Q_f \gamma^\mu\ .
  \label{eq:kinetic:mix:fermion:coupling}
\end{align}
We thus refer to recent reviews to summarize those bounds~\cite{Battaglieri:2017aum,Alexander:2016aln}. It is sufficient to note that for the range of dark photon masses of interest there is always a sufficiently small $\varepsilon$ (large $\Lambda$) such that the basic phenomenology is valid. In the small mixing limit, mediators are very long lived and may be targets for recently proposed indirect detection techniques~\cite{Rothstein:2009pm, Slatyer:2016qyl, Kim:2017qaw, Gori:2018lem}

To demonstrate the phenomenology of kinetic mixing, we examine the bounds on our scenario coming from direct detection experiments. The $W$--nucleon ($N$) scattering amplitude, $i\mathcal M_N$, is
\begin{align}
  \vcenter{
    \hbox{\includegraphics[width=.2\textwidth]{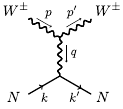}}
    }
  &=
  \frac{ g \varepsilon e Q_N}{q^2-m_{A}^2}\varepsilon_{\mu}(p)\varepsilon_{\nu}^*(p')\bar{u}(k')\left[g^{\mu \nu}\left(\slashed{p}+\slashed{p'}\right)-2p'^{\mu}\gamma^{\nu}-2p^{\nu}\gamma^{\mu}\right]u(k) \label{eq:Mnrelativistic}
  \ ,
\end{align}
where we recall that here $g$ is the dark sector gauge coupling in (\ref{eq:darkphotonDMrule}--\ref{eq:darkphotonDMcontactrule}). This interaction maps simply to the non-relativistic $\mathcal O^{(\text{NR})}_1$ (spin-independent) operator~\cite{Fan:2010gt,Fornengo:2011sz,DelNobile:2013sia,Dent:2015zpa,Catena:2018uae,Catena:2019hzw}. In the notation of Refs.~\cite{Dent:2015zpa, Catena:2019hzw}, the interactions \eqref{eq:darkphotonDMrule} and \eqref{eq:kinetic:mix:fermion:coupling} map onto effective couplings $b_5 = g$ and $h_3 = \varepsilon e Q_q$ so that the non-relativistic effectivie coupling to nucleons is
\begin{align}
  c_1^N &\equiv - \frac{b_5 h_3^N}{m_A^2}
  &\text{from which we define}&&
  c_p \equiv |c_1^N| = \frac{\varepsilon e g}{m_A^2} \ .
\end{align}
We have used the fact that the dark photon coupling to nucleons, $h^N$, is proportional to the sum of the valence charges of the quarks due to the conservation of the electric current.

\begin{figure}[t]
\begin{center}
	\includegraphics[width=\textwidth]{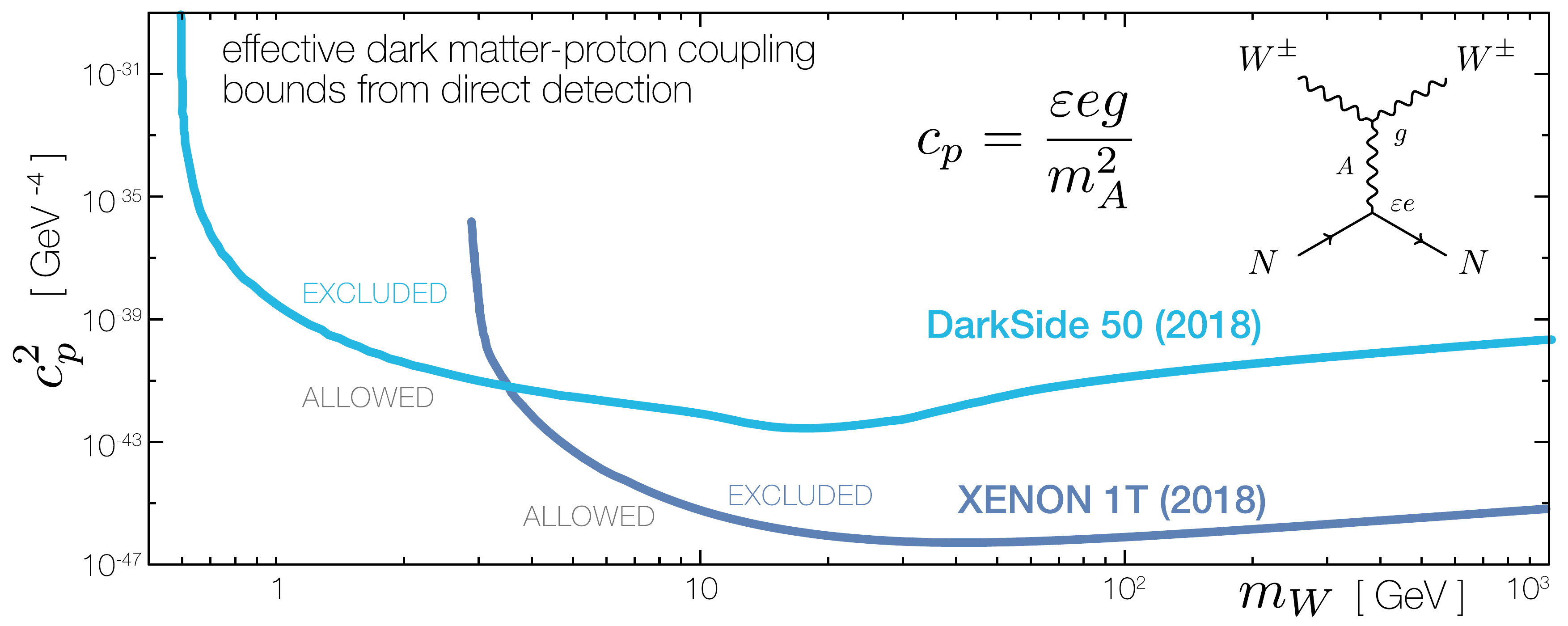}
\end{center}
\caption{Constraints on the effective dark matter--proton coupling, $c_p^2$, from direct detection experiments  \acro{XENON~1T}~\cite{Aprile:2018dbl} and DarkSide 50~\cite{Agnes:2018fwg}.}
\label{fig:DDbounds}
\end{figure}

We compare the effective coupling $c_p$ to the most stringent bounds on spin-independent dark matter--nucleon scattering: \acro{XENON~1T}~\cite{Aprile:2018dbl} and DarkSide 50~\cite{Agnes:2018fwg}. The results are presented in Fig.~\ref{fig:DDbounds}. For a given mediator mass $m_A$ and dark gauge coupling $g$, this sets an effective bound on the size of the kinetic mixing parameter $\varepsilon$ or, alternatively, the effective scale $\Lambda$ of the dimension-5 operator, \eqref{eq:kin:mix:dim:5}. For very small values of $\varepsilon$ one may realize unique thermal histories that are beyond the scope of this study~\cite{Chu:2011be,Blennow:2013jba,Krnjaic:2017tio,Evans:2017kti, Dvorkin:2019zdi}.
We remark that in the event of a discovery of dark matter scattering at direct detection experiments, Refs.~\cite{Dent:2015zpa, Catena:2019hzw} show that vector dark matter interacting through a vector mediator may be disentangled from other candidate models through its recoil spectrum.

\section{Conclusions}

This manuscript presents the first model of a stable, vector dark matter with a low-mass vector mediator.
We present a full theory with the required scalar sector to enact the necessary symmetry breaking pattern and explain the stabilization mechanism from symmetry principles. This model can be understood as the Higgsed phase of a Yang--Mills hidden sector, in contrast to the confined glueball-dark matter phase explored in Refs.~\cite{Boddy:2014qxa,Boddy:2014yra}. We present the basic phenomenology assuming that the dark matter abundance is produced by thermal freeze out. We present benchmark parameters for self-interacting dark matter solutions to small scale astrophysical anomalies where we observe a slight tension between the parameters required for a thermal relic and those that can fit the inferred self-interaction cross section across a range of systems from dwarfs to clusters.
We leave detailed self-interacting dark matter fits for small scale structure anomalies to future work as this is likely to require additional model building to navigate cosmological bounds and abundance~\cite{Huo:2017vef}.
We also present bounds from direct detection assuming that the vector mediator is the primary portal to the Standard Model, in contrast to similar theories of vector dark matter that assumed a Higgs portal.

Our model is a minimal framework for a spin-1 dark sector that can be mapped on to standard dark sector phenomenology.
This model offers many directions for further exploration. Within the perturbative regime of this theory, we identified possibilities for producing the dark matter abundance beyond the thermal freeze out assumption. This connects to recent and ongoing work on the phase space of dark sectors whcih thermalize through a portal interaction~\cite{Feng:2008mu, Chu:2011be,Blennow:2013jba,Krnjaic:2017tio,Evans:2017kti, Dvorkin:2019zdi}. Depending on the production mechanism, dark matter may be symmetric or asymmetric, which in turn feeds into the self-interaction phenomenology by affecting the possibility of self-interaction resonances. The model contains an additional light mediator ($H_1$) and an additional charged particle ($\pi^\pm$) that we assumed to be negligible in this work. One can imagine an interplay of the two mediators  for $t$-channel processes such as self-interactions or direct detection, or alternatively inelasticity coming from a small splitting between the $\pi^\pm$ and $W^\pm$. One may alternatively push $m_W > m_\pi$ so that the stable dark matter candidates are charged scalars with derivative interactions to a dark photon. Finally, we remark that in the non-perturbative regime the model also furnishes dark sector 't~Hooft--Polyakov monopoles~\cite{Baek:2013dwa, Khoze:2014woa}. Our model simply realizes the regime where the Abelian force associated with the monopoles is Higgsed, therefore the monopoles are expected to confine. This, in turn, may new dynamics relevant for the dark sector~\cite{Terning:2018udc,Terning:2018lsv}.

\section*{Acknowledgments}

We are supported by the \acro{DOE} grant \textsc{de-sc}/0008541.
We thank Hai-Bo Yu and Gerardo Alvarez for thoughtful discussions about self-interacting dark matter and for providing data from Fig.~1 of Ref.~\cite{Kaplinghat:2015aga}. We thank Jose Wudka for insights that helped us identify the \UU{(1)}$_{H'}$ symmetry that stabilizes the $W$. We acknowledge work by Oleg Popov and Corey Kownacki on an early version of this project. We appreciate correspondence with James Dent where he shared an advanced version of errata for \cite{Dent:2015zpa} that we used for our direct detection analysis. We appreciate feedback from Bhaskar Dutta, William Shepherd, Jonathan Feng, Timothy M.P.~Tait, and Aniket Joglekar at various stages of this work.

\textsc{p.t.}~thanks the Aspen Center for Physics (\acro{NSF} grant \#1066293) and the Kavli Institute for Theoretical Physics (\acro{NSF} \acro{PHY}-1748958) for their hospitality during periods where part of this work was completed. We also thank Brian Shuve and Harvey Mudd College for hosting us as we were  completing this work.

We used the \textit{Mathematica} computer algebra system for calculations and plotting~\cite{Mathematica}; we used the \textit{FeynCalc} package for cross section computations ~\cite{Shtabovenko:2016sxi}. We used \textit{Affinity Designer} for post-production of figures and plots\footnote{We are flattered that so many of our colleagues keep asking. And yes, there is a slight learning curve.}~\cite{AffinityDesigner} with the Apple mac{os} font Helvetica Neue.

\appendix

\section{Goldstones and Pions: an Abelian Example}
\label{app:U1:example}

We present a simple model to demonstrate the parameterization of the Goldstone degrees of freedom in \eqref{eq:vectorandaxialstates} and some of the nuances in the discussion of Section~\ref{eq:vectorandaxialstates}. Let $a(x)$ and $b(x)$ be complex scalar fields with potentials such that $\langle a(x) \rangle = f_a/\sqrt{2}$ and $\langle b(x)\rangle = f_b/\sqrt{2}$. We pass to a non-linear representation,
\begin{align}
  a(x) &= \frac{r_a(x)}{\sqrt{2}} \, e^{i\varphi_a(x)/f_a}
  &
  b(x) &= \frac{r_b(x)}{\sqrt{2}} \, e^{i\varphi_b(x)/f_b} \ .
\label{eq:a:b:U(1):example}
\end{align}
The vevs $f_a$ and $f_b$ are order parameters for the breaking patterns
\begin{align}
  \UU{(1)}_a &\to \varnothing
  &
  \UU{(1)}_b &\to \varnothing \ ,
\end{align}
where $\UU{(1)}_{a,b}$ correspond to separate rephasing of the $a$ and $b$ fields. Focusing on the Goldstone degrees of freedom, we may take $r_i(x)\to f_i$. If the Lagrangian respects the $\UU{(1)}_{a}\times\UU{(1)}_{b}$ symmetry then the Goldstone fields are independent, massless, free degrees of freedom.

\subsection{Gauging a Subgroup Combination}

Suppose we gauge a subgroup of $\UU{(1)}_a\times \UU{(1)}_b$ under which the $A$ and $B$ fields have charges $q_a$ and $q_b$ respectively. The covariant derivative of this gauged symmetry is
\begin{align}
  D_\mu = \partial_\mu - i g q_i W_\mu \ ,
\end{align}
where $g$ is the gauge coupling and $W_\mu$ is the gauge boson. Ignoring the radial excitations, the kinetic terms for $a$ and $b$ yield
\begin{align}
  |Da|^2 + |Db|^2 & =
  \frac 12 (\partial \varphi_a)^2
  +
  \frac 12 (\partial \varphi_b)^2
  -
  g\partial\left(q_a f_a \varphi_a + q_b f_b \varphi_b\right)\cdot W
  +
  \frac{g^2}{2}\left(q_a^2f_a^2+q_b^2f_b^2\right) W^2 \ .
  \label{eq:U(1):example:kinetic}
\end{align}
We see that the gauge boson $W$ picks up a mass and eats a linear combination of the Goldstone bosons. We identify the effective order parameter $f_W$ for the gauge symmetry breaking and the mass of the $W_\mu$:
\begin{align}
  f_W^2 &= q_a^2f_a^2 + q_b^2f_b^2
  &
  m_W^2 = g^2 f_W^2 \ .
\end{align}
The Goldstone combination that is eaten, $\varphi_W(x)$, and its orthogonal combination $\varphi_X(x)$ are
\begin{align}
  \varphi_W
  &=
  \frac{q_a f_a}{f_W}\varphi_a + \frac{q_b f_b}{f_W}\varphi_b
  &
  \varphi_X
  &=
    \frac{q_b f_b}{f_W} \varphi_a
    -
    \frac{q_a f_a}{f_W}
    \varphi_b
  \ .
\end{align}
Observe that the eaten Goldstone is mostly composed of the field which contributes more to the gauge symmetry breaking. Thus if $q_a f_a > q_b f_b$, then $\varphi_W$ contains more of the $a$ phase than the $b$ phase. The orthogonal combination, $\varphi_X$, is a bona-fide Goldstone mode in the theory and is composed of mostly the phase of the subdominant source of symmetry breaking.

\subsection{Gauging a Vectorlike Combination}

Suppose that we gauge $\UU{(1)}_V$, the diagonal subgroup of $\UU{(1)}_a\times \UU{(1)}_b$. Under $\UU{(1)}_V$, both $a$ and $b$ have the same charge, $q_a = q_b=1$. The analysis above yields
\begin{align}
  f_V^2 &= f_a^2 + f_b^2
  &
  \varphi_V &= \frac{f_a}{f_V}\varphi_a + \frac{f_b}{f_V}\varphi_b
  &
  \varphi_A &= \frac{f_b}{f_V}\varphi_a - \frac{f_a}{f_V}\varphi_b
  \ .
  \label{eq:U(1):vec}
\end{align}
Here $\varphi_A$ is the Goldstone for the `axial' rotation under which $a$ and $b$ transform with opposite phase, $q_a = - q_b = 1$, and is orthogonal to the Goldstone for the vector rotation $\varphi_V$ that is eaten by $V_\mu$.
This is analogous to the case of electroweak symmetry breaking where the Higgs order parameter for $\SU{(2)}_L\times\UU{(1)}_Y$ is much larger than that of the \acro{QCD} chiral condensate so that the longitudinal modes of massive electroweak bosons are mostly components of the Higgs doublet. The pions are [pseudo-]Goldstone bosons analogous to the $\varphi_A$: they are mostly composed of the phase of the chiral condensate, but contain a small piece of the Higgs doublet that is shifted by the opposite symmetry transformation parameter.

\subsection{Which Goldstone is Which?}

\begin{figure}
  \begin{center}
    \includegraphics[width=\textwidth]{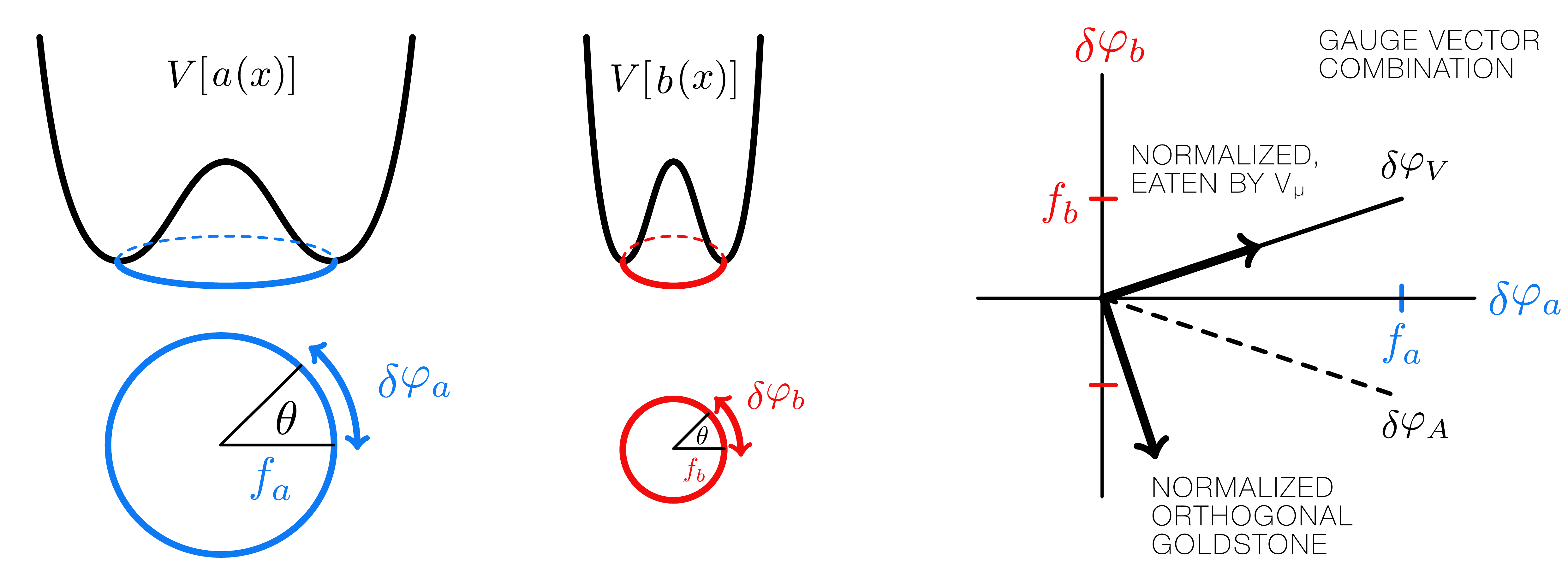}
  \end{center}
  \caption{Fields $a$ and $b$ acquire unequal vacuum expectation values $f_a > f_b$. The Goldstone excitations with respect to a transformation by parameter $\theta$ have correspondingly different magnitudes, $\delta\varphi_a > \delta \varphi_b$. The Goldstone, $\delta \varphi_V$, for a vectorial transformation where $\theta_a = \theta_b$ is thus \emph{not} orthogonal to the corresponding Goldstone, $\delta\varphi_A$ for an axial transformation where $\theta_a = -\theta_b$.}
  \label{fig:Goldstones}
\end{figure}

This presents a puzzle. One is free to describe the symmetry structure of the theory with respect to $\UU{(1)}_a\times \UU{(1)}_b$ or $\UU{(1)}_V\times \UU{(1)}_A$.
Suppose $f_a \gg f_b$. Then in the $\UU{(1)}_a\times \UU{(1)}_b$ description, the $a$ field plays a bigger role in symmetry breaking than the $b$ field. However, the vevs each break $\UU{(1)}_V$ and $\UU{(1)}_A$ by the same effective order parameter, $f_V^2 = f_a^2 + f_b^2$. Neither $\UU{(1)}_V$ nor $\UU{(1)}_A$ is preferred over the other. Why, then, is it the case in \eqref{eq:U(1):vec} that the $\varphi_V$ eats more $\varphi_a$ while $\varphi_A$ eats more of $\varphi_b$? The root of this confusion is illustrated in Fig.~\ref{fig:Goldstones}: in the absence of gauging, the na\"ive description of the vector and axial Goldstones are not orthogonal to one another. The choice of gauging a particular combination of the full global symmetry breaks the symmetry and gives `priority' to the eaten Goldstone boson to have a larger admixture of the field that does most of the symmetry breaking.

\subsection{Gauging an Axial Combination}

One way to illustrate this point is to observe that if we had instead gauged the axial symmetry, $q_a = -q_b = -1$. Let us continue to assume that $f_a\gg f_b$. The order parameter for axial symmetry breaking is identical to the vector case so that the axial symmetry, $f_A = f_V$. The only difference from the vector case is that it is now the axial Goldstone, $\varphi_A$ that is eaten:
\begin{align}
  f_A^2 &= f_a^2 + f_b^2
  &
  \varphi_A &= \frac{f_a}{f_V}\varphi_a - \frac{f_b}{f_V}\varphi_b
  &
  \varphi_B &= \frac{f_b}{f_V}\varphi_a + \frac{f_a}{f_V}\varphi_b
  \ .
  \label{eq:U(1):axial}
\end{align}
Observe that compared to \eqref{eq:U(1):vec}, the relative admixtures of $\varphi_{a,b}$ has changed so that the eaten Goldstone (now $\varphi_A$) is \emph{still} mostly composed of the Goldstone ($\varphi_a$) from the dominant source of symmetry breaking. This follows directly from \eqref{eq:U(1):example:kinetic} where it is clear that the choice of which symmetry is gauged determines which linear combination of fields has more of the $\varphi_a$ field.

\subsection{Gauging both Vector and Axial Symmetry}

Another illustrative example is to separately gauge the vector and axial combinations with gauge couplings $g_V$ and $g_A$ respectively. We are primarily interested in the case $g_V = g_A$, but the two are independent parameters. The kinetic terms then include
\begin{align}
  |Da|^2 + |Db|^2 &\supset
  \partial \varphi_a\cdot(g_VV + g_AA) + \partial \varphi_b\cdot (g_VV - g_A A)
  +
  \frac{f_a^2}2 (g_VV + g_AA)^2
  +
  \frac{f_b^2}2 (g_VV - g_AA)^2 \ .
\end{align}
In this case the gauge boson mass matrix is not diagonal. When $g_V = g_A$ this matrix is diagonalized by writing
\begin{align}
  V &= W+Z
  &
  A & = W-Z \ .
\end{align}
This transformation is independent of the relative magnitudes of the vevs. The transformation also separates the mixing terms:
\begin{align}
  g\partial \varphi_a\cdot(V + A) + g\partial \varphi_b\cdot (V - A)
  =
  g\partial \varphi_a\cdot W + g\partial \varphi_b\cdot Z \ .
\end{align}
Thus we are pushed back to the natural basis of Goldstone bosons, $\varphi_{a,b}$. The vector and axial gauge bosons are forced to mix in such a way that the mass eigenstates end up being a gauge boson that eats $\varphi_a$ and a gauge boson that eats $\varphi_b$. This is equivalent to the case where one separately gauges the $\UU{(1)}_a$ and $\UU{(1)}_b$ symmetries.

\subsection{Global Vector and Axial Goldstones}

As a final exercise, one may consider the ungauged theory where one writes the fields in terms of axial and vector Goldstones. In \eqref{eq:a:b:U(1):example} one would then identify
\begin{align}
  \varphi_a &= \frac{f_a}{\sqrt{2f_a^2 + 2f_b^2}} \left(\varphi_V + \varphi_A\right)
  &
  \varphi_b &= \frac{f_b}{\sqrt{2f_a^2 + 2f_b^2}} \left(\varphi_V - \varphi_A\right)
  \ ,
\end{align}
where the normalizations are chosen so that (1) an excitation along, say, the $\varphi_V$ direction produces an equal phase transformation on $a(x)$ and $b(x)$ and (2) the $\varphi_{V,A}(x)$ are canonically normalized. Here we see that in the absence of gauging, the $\varphi_{V,A}$ are treated `equally' despite the unequal vevs. The scenario is identical to the description in terms of $\varphi_{a,b}$ in that the fields are massless, free excitations.

\section{SIDM Methodology}
\label{app:SIDM:method}

We summarize the methodology for determining the dark matter self-interaction cross section as a function of velocity in Fig.~\ref{fig:SIDMplot}. We closely follow the procedure in Ref.~\cite{Tulin:2013teo}.
The relevant quantity is the transfer cross section,
\begin{equation}
\sigma_T=\int d\Omega \left(1-\cos\theta\right)\frac{d \sigma}{d \Omega} \ , \label{eq:transfercrosssection}
\end{equation}
which characterizes interaction cross section weighted by momentum transfer. This regulates the $\cos\theta \to 1$ divergence where dark matter scatters do not affect halo shapes. There is no known analytical expression for the transfer cross section that valid for the entire parameter space though it has been calculated under various approximations for limited parts of parameter space~\cite{Feng:2009hw,Ibe:2009mk,Loeb:2010gj,Lin:2011gj,Aarssen:2012fx,Buckley:2009in}. A large part of the parameter space corresponds to the resonant regime where both quantum mechanical and non-perturbative effects become important, as such a numerical solution to the non-relativistic Schr\"odinger equation is necessary.

We use a partial wave analysis. The transfer cross section is related to the $\ell^{\text{th}}$ partial wave phase shift, $\delta_\ell$, by
\begin{equation}
\sigma_T=\frac{4 \pi}{k^2}\sum\limits_{\ell=0}^{\infty}(\ell+1)\sin^2\left(\delta_{\ell+1}-\delta_{\ell}\right) \label{eq:transfercrosssectionsum} \ .
\end{equation}
The $\delta_{\ell}$s are, in turn, obtained by solving the radial Schr\"{o}dinger equation
\begin{equation}
\frac{1}{r^2}\frac{d}{dr}\left(r^2\frac{dR_{\ell}}{dr}\right)+\left(k^2-\frac{\ell(\ell+1)}{r^2}-m_X V(r)\right)R_{\ell}=0 \label{eq:Schroeq} \ ,
\end{equation}
where $k=m_X v/2$ and $v$ is the relative velocity of the two-particle dark matter system. $\delta_l$ is found by comparing with the asymptotic solution for $R_{\ell}$:
\begin{equation}
\lim\limits_{r\to \infty}R_{\ell}(r) \propto \cos \delta_{\ell} j_{\ell}(kr)-\sin \delta_{\ell} n_{\ell}(kr) \label{eq:asympsol}\ ,
\end{equation}
where $j_{\ell}$ ($n_\ell$) is the spherical Bessel (Neumann) function of the $\ell^{\text{th}}$ kind.
We define the function $\chi_{\ell} \equiv r R_{\ell}$ and dimensionless variables
\begin{align}
x &\equiv \alpha_X m_X r & a&=\frac{v}{2\alpha_X} & b&=\frac{\alpha_X m_X}{m_{\phi}} \ ,
\end{align}
  so that we can rewrite \eqref{eq:Schroeq} as~\cite{Buckley:2009in}
\begin{equation}
\left(\frac{d^2}{dx^2}+a^2-\frac{\ell (\ell+1)}{x^2}\pm \frac{1}{x}e^{-x/b}\right)\chi_{\ell}(x)=0\ . \label{eq:dimlessSchroeq}
\end{equation}
Near the origin, the non-derivative parts of \eqref{eq:dimlessSchroeq} are dominated by the angular momentum term. This implies that $\chi_{\ell}\propto x^{\ell+1}$ close to $x=0$. We choose a normalization such that $\chi_{\ell}(x_0)=1$ and $\chi'_{\ell}(x_0)=\left(\ell+1\right)/x_0$ where $x_0$ is a point close to the origin chosen to satisfy $x_0 \ll b$ and $x_0 \ll \left(\ell+1\right)/a$. We use $x_0$ as the lower limit for range in which we numerically solve the Schr\"odinger equation. Similarly, to define the upper limit of range, we pick a point $x_m$ satisfying the condition $a^2 \gg \exp\left(-x_m/b\right)/x_m$. When $x_m$ satisfies this condition, the potential term is neglible compared to the kinetic term and the solution approaches
\begin{equation}
\chi_{\ell}(x)\propto x e^{i \delta_{\ell}}\left(\cos\delta_{\ell}j_{\ell}(a x)-\sin\delta_{\ell} n_{\ell}(a x)\right).
\end{equation}
The phase shift is then
\begin{align}
\tan \delta_{\ell}&=\frac{a x_m j'_{\ell}(a x_m)-\beta_{\ell}j_{\ell}(a x_m)}{a x_m n'_{\ell}(a x_m)-\beta_{\ell}n_{\ell}(a x_m)}
& \text{where}&&
\beta_{\ell}&=\frac{x_m \chi'_{\ell}(x_m)}{\chi_{\ell}(x_m)}-1 \ .
\label{eq:tandelta}
\end{align}
For an initial guess of the range $(x_0,x_m)$ and the maximum number of partial waves required for convergence, $\ell_\text{max}$, we calculate $\delta_{\ell}$ from \eqref{eq:tandelta}. We then increase $x_m$ and decrease $x_0$, recalculating $\delta_{\ell}$ until the differences of successive iterations converge to be within 1\%. We then sum \eqref{eq:transfercrosssectionsum} from $\ell= 0$ to $\ell=\ell_{\text{max}}$ to obtain an estimate for $\sigma_T$.
Next we increment $\ell_{\text{max}}\to \ell_\text{max}+1$ and repeat the procedure until successive values of $\sigma_T$ converge to be within 1\% and $\delta_{\ell_{\text{max}}}<0.01$.
Ref.~\cite{Tulin:2013teo} iterates $\ell_{\text{max}}$ until $\sigma_T$ converged and $\delta_{\ell_{\text{max}}}<0.01$ ten consecutive times. We have found that for our analysis that it is sufficient to stop the calculation after one successful convergence. We have found that the ``StiffenessSwitching'' method from the \texttt{NDSolveUtilities} package in \emph{Mathematica} to be particularly useful.

Our model exhibits both attractive and repulsive self-interactions due to dark matter being symmetric and mediated by a vector particle. In this case, one solves the Schr\"odinger equation separately for each sign of the potential to extract two transfer cross sections, $\sigma_T^{(\pm)}$. The effective transfer cross section is the average of the two.

\bibliographystyle{utcaps} 	
\bibliography{VectorSIDM}

\end{document}